\documentclass[a4paper,superscriptaddress,showpacs,showkeys,twocolumn]{revtex4-1}
\pdfoutput=1
\usepackage{amsmath,amsfonts,amssymb,bm,color,graphicx}
\usepackage{epsfig,grffile,tipa}
\usepackage{subcaption}

\renewcommand{\vec}[1]{\bm{#1}}
\renewcommand{\tensor}[1]{\vec{\mathsf{#1}}}
\newcommand{\stensor}[1]{\mathsf{#1}}
\newcommand{\uvec}[1]{\hat{\bm{#1}}}

\newcommand{\vbody}[1]{\widetilde{\bm{#1}}}
\newcommand{\df}[1]{\text{d}#1\,}
\newcommand{\vdf}[1]{\text{d}\vec{#1}\,}

\newcommand{\norm}[1]{\left\lVert#1\right\lVert}
\newcommand{\fnorm}[1]{\left\lVert#1\right\lVert_F}

\DeclareGraphicsExtensions{{.pdf,.eps,.png}}
\begin{document}
 \title{Direct numerical simulations of rigid body dispersions. I.
Mobility/Friction tensors of assemblies of spheres}
 \author{John J. Molina} \email{john@cheme.kyoto-u.ac.jp}
 \affiliation{Department of Chemical Engineering, Kyoto University,
   Kyoto 615-8510}
 \author{Ryoichi Yamamoto} \email{ryoichi@cheme.kyoto-u.ac.jp}
 \affiliation{Department of Chemical Engineering, Kyoto University,
   Kyoto 615-8510} \date{\today}
 \begin{abstract}
   An improved formulation of the ``Smoothed Profile'' method is
   introduced to perform direct numerical simulations of arbitrary
   rigid body dispersions in a Newtonian host solvent. Previous
   implementations of the method were restricted to spherical
   particles, severely limiting the types of systems that could be
   studied. The validity of the method is carefully examined by
   computing the friction/mobility tensors for a wide variety of
   geometries and comparing them to reference values obtained from
   accurate solutions to the Stokes-Equation.
 \end{abstract}
 \pacs{}
 \keywords{hydrodynamic interactions, direct numerical simulations,
   friction, mobility}
\maketitle
\section{Introduction}
The dynamics of particles dispersed in a host solvent, how they react
to and affect the fluid motion, is a relevant problem for science as
well as engineering fields\cite{Russel:1992wr}. It is necessary to
account for the macroscopic properties of suspensions (such as the
viscosity, elastic modulus, and thermal conductivity), as well as the
mechanics of protein unfolding\cite{Yamakawa:1970dv}, the kinetics of
bio-molecular reactions\cite{Son:2011bo}, and the tumbling motion of
bacteria\cite{Sokolov:2012td}. As the particles move, they generate
long-range disturbances in the fluid, which are transmitted to all
other particles. Properly accounting for these hydrodynamic interactions
has proven to be a very complicated task due to their non-linear
many-body nature. 

Several numerical methods have been developed to explicitly include
the effect of hydrodynamic interactions in a suspension of
particles\cite{BRADY:1988up,Cichocki:1994gl,Malevanets:1999ki,Hu:2001hq,
  Glowinski:2001bf,LADD:2001tt,Cates:2004kh}.
However, their applicability to non-Newtonian host solvents or
solvents with internal degrees of freedom is not straightforward, and
in some cases not possible. We have proposed an alternative direct
numerical simulation (DNS) method, which we refer to as the Smooth
Profile (SP) method, that simultaneously solves for the host fluid and
the particles. The coupling between the two motions is achieved
through a smooth profile for the particle interfaces. This method is
similar in spirit to the fluid particle dynamics
method\cite{Tanaka:2000gq}, in which the particles are modeled as a
highly viscous fluid. The main benefit of our model is the ability to
use a fixed cartesian grid to solve the fluid equations of motion. 

The SP method has been successfully used to study the diffusion,
sedimentation, and rheology of colloidal dispersions in incompressible
fluids\cite{Iwashita2009b,Kobayashi2010a,Iwashita2009a,Kobayashi:2011tm,Hamid:2013bo}.
Recently it has been extended to include self-propelled
swimmers\cite{Molina:2013hq} and for compressible host
solvent\cite{Tatsumi:2012ek}. So far, however, only spherical
particles were considered. In this paper we extend the SP method to be
applicable to arbitrary rigid bodies. We show the validity of the
method by computing the mobility/friction tensors for a large variety
of geometric shapes. The results are compared to numerical solutions
of the Stokes equation, which are essentially exact, as well as
experimental data. The agreement with our results is excellent in all
cases considered here. Future papers in this series will deal with the
dynamical properties of rigid body dispersions in detail, with this
work we aim to introduce the basics of the model and show its
validity.

\section{Simulation Model}
\subsection{Smooth Profile Method for Rigid Bodies}
We solve the dynamics of a rigid body in an incompressible Newtonian
host fluid using the SP method. The basic idea behind this method is
to replace the sharp boundaries between the particles and the fluid
with a smooth interface. This allows us to define all field variables
over the entire computational domain, and results in an efficient
method to accurately resolve the many-body hydrodynamic interactions. 

The motion of the host fluid is governed by the incompressible
Navier-Stokes equation\cite{Batchelor:2000uu}
\begin{align}
  \left(\partial_t + \vec{u}_f\cdot\nabla\right)\vec{u}_f &=
  \rho_f^{-1}\nabla\cdot\tensor{\sigma}_f\label{e:ns_host}\\
  \nabla\cdot \vec{u}_f &= 0
\end{align}
where $\vec{u}_f$ is the fluid velocity, $\rho_f$ the density, and
$\tensor{\sigma}_f$ the Newtonian stress tensor
\begin{align}
  \tensor{\sigma}_f &= -p\tensor{I} + \tensor{\sigma}^\prime \label{e:stress}\\
  \tensor{\sigma}_f^\prime &= \eta_f\left[\nabla\vec{u}_f +
    \left(\nabla\vec{u}_f\right)^t\right]\label{e:dstress}
\end{align}
with $p_f$ and $\eta_f$ the pressure and viscosity of the fluid,
respectively. 

The motion of the particle is given by the Newton-Euler
equations\cite{Jose:1998vq},
\begin{align}
  \dot{\vec{R}}_I &= \vec{V}_I &\dot{\tensor{Q}}_I &=
  \text{skew}\left(\vec{\Omega}_I\right)\cdot\tensor{Q}_i \label{e:ne_pos}\\
  M_I\dot{\vec{V}}_I &= \vec{F}_I & \tensor{I}_I\cdot\vec{\Omega}_I&=
  \vec{N}_I\label{e:ne_vel}
\end{align}
where $\vec{R}_I$, $\vec{V}_I$, $\vec{\Omega}_I$, and $\tensor{Q}_I$,
are the center of mass (com) position, velocity, angular velocity, and
orientation matrix, respectively, of the $I$-th rigid body
($I=1,\cdots,N$). The total force and torque experienced the particles
is denoted as $\vec{F}_I$ and $\vec{N}_I$, respectively, with
$\tensor{I}_I$ the moment of inertia tensor and
$\text{skew}(\vec{\Omega}_I)$ the skew-symmetric angular velocity
matrix
\begin{align}\label{e:skew}
  \text{skew}\left(\vec{\Omega}\right) = \begin{pmatrix}
    0 & -\Omega^z & \Omega^y \\
    \Omega^z & 0 & -\Omega^x \\
    -\Omega^y & \Omega^x & 0
  \end{pmatrix}
\end{align}
The force (and torque) on each of the particles
$\vec{F}_I=\vec{F}_I^{\text{H}} + \vec{F}_I^{\text{C}} +
\vec{F}_I^{\text{Ext}}$ is comprised of hydrodynamic contributions
$\vec{F}^{\text{H}}$, particle-particle interactions (including a core
potential to prevent overlap) $\vec{F}^{\text{C}}$, and a possible
external field contribution $\vec{F}^{\text{Ext}}$. For the present
study, we neglect thermal fluctuations, but they are easily included
within the SP formalism\cite{Iwashita:2008cj}.

The coupling between fluid and particles is obtained by defining a
total velocity field $\vec{u}$, with respect to the fluid $\vec{u}_f$
and particle $\vec{u}_p$ velocity fields, as
\begin{align}
  \vec{u}(\vec{x}) &= \left(1 - \phi\right)\vec{u}_f(\vec{x}) +
  \phi\vec{u}_p(\vec{x}) \label{e:u} \\
  \phi\vec{u}_p(\vec{x}) &= \sum_I\phi_I(\bm{x};\vec{R}_I,\tensor{Q}_I)\left[\vec{V}_I +
    \vec{\Omega}_I\times\vec{r}_I\right] \label{e:up}
\end{align}
where $\phi_I(\bm{x};\bm{R}_I, \tensor{Q}_I)$ is a suitably defined SP
function ($0\le \phi_I \le 1$) that interpolates between fluid and
particle domains (as described below), and $\vec{r}_I=\vec{x} -
\vec{R}_I$. The modified Navier-Stokes equation which governs the
evolution the total fluid velocity field (host fluid $+$ particle) is
given by\cite{Nakayama2005a}
\begin{align}
  \left(\partial_t + \vec{u}\cdot\nabla\right)\vec{u} &= \rho^{-1}
  \nabla\cdot\tensor{\sigma}
  + \phi\vec{f}_p \label{e:ns}\\
  \nabla\cdot\vec{u}_f &= 0
\end{align}
with $\rho=\rho_f$ and $\eta=\eta_f$. The stress tensor is defined as
in eq.~\eqref{e:stress}, but in terms of the total fluid velocity
$\vec{u}$.

The scheme used to solve the equations of motion is the same
fractional-step algorithm introduced in ref\cite{Nakayama:2008fi}.,
with minor modifications needed to account for the non-spherical
geometry of the particles. Let $\vec{u}^n$ be the velocity field at
time $t_n=n h$ ($h$ the time interval). We first solve for the
advection and hydrodynamic viscous stress terms, and propagate the
particle positions (orientations) using the current particle
velocities
\begin{align}
  \vec{u}^* &= \vec{u}^n +
  \int_{t_n}^{t_n+h}\df{s}\nabla\cdot\left[\rho^{-1}\tensor{\sigma} -
    \vec{u}\vec{u}\right] \label{e:ustar}\\
  \vec{R}_I^{n+1} &= \vec{R}_I^n + \int_{t_n}^{t_n+h}\df{s}\vec{V}_I \label{e:rn}
  \\
  \tensor{Q}_I^{n+1} &= \tensor{Q}_I^{n} +
  \int_{t_n}^{t_n+h}\df{s}\text{skew}\left(\vec{\Omega}_I\right)\cdot\tensor{Q}_I\label{e:qn}
\end{align}
Given the dependence of the profile function on the particle position
and orientation, we must also update the particle velocity field to
\begin{align}
  \phi\vec{u}_p^*(\vec{x}) = \sum_I
  \phi^{n+1}_I(\vec{x})\left[\vec{V}_I^{n+1} + \vec{\Omega}_I\times\vec{r}_I^{n+1}\right]
\end{align}

Next, we compute the hydrodynamic force and torque exerted by the
fluid on the particles, by assuming momentum conservation. The time
integrated hydrodynamic force and torque over a period $h$ are equal
to the momentum exchange over the particle domain
\begin{align}
  \left[\int_{t_n}^{t_n+h}\df{s}\vec{F}_I^{\text{H}}\right] &=
  \int\vdf{x}\rho\phi_I^{n+1}\left(\vec{u}^* - \vec{u}_p^{*}\right) \\
  \left[\int_{t_n}^{t_n+h}\df{s}\vec{N}_I^{\text{H}}\right] &=
  \int\vdf{x}\left[\vec{r}_I^{n+1}\times
    \rho\phi_I^{n+1}\left(\vec{u}^* - \vec{u}_p^*\right)\right]
\end{align}
From this, and any other forces on the rigid bodies, we update the
velocities of the particles as
\begin{align}
  \vec{V}_I^{n+1} &= \vec{V}_I^{n} + 
  M_I^{-1}\int_{t_n}^{t_n+h}\df{s}\left[\vec{F}_I^{\text{H}} + \vec{F}_I^{\text{C}} +
    \vec{F}_I^{\text{Ext}}\right] \\
  \vec{\Omega}_I^{n+1} &= \vec{\Omega}_I^{n} + 
  \tensor{I}_I^{-1}\int_{t_n}^{t_n+h}\cdot\left[\vec{N}_I^{\text{H}} +
    \vec{N}_I^{\text{C}} + \vec{N}_I^{\text{Ext}}\right]
\end{align}
Finally, the particle rigidity is imposed on the total fluid velocity
through the body force in the Navier-Stokes equation
\begin{align}
  \vec{u}^{n+1} &= \vec{u}^* +
  \left[\int_{t_n}^{t_n+h}\df{s}\phi\vec{f}_p\right] \\
  \left[\int_{t_n}^{t_n+h}\df{s}\phi\vec{f}_p\right] &=
  \phi^{n+1}\left(\vec{u}_p^{n+1} - \vec{u}^{*}\right) -
  \frac{h}{\rho}\nabla p_p
\end{align}
with the pressure due to the rigidity constraint obtained from the
incompressibility condition $\nabla\cdot\vec{u}^{n+1} = 0$. 
\subsection{Rigid Body Representation}
\begin{figure}
  \centering
  \includegraphics[width=0.35\textwidth]{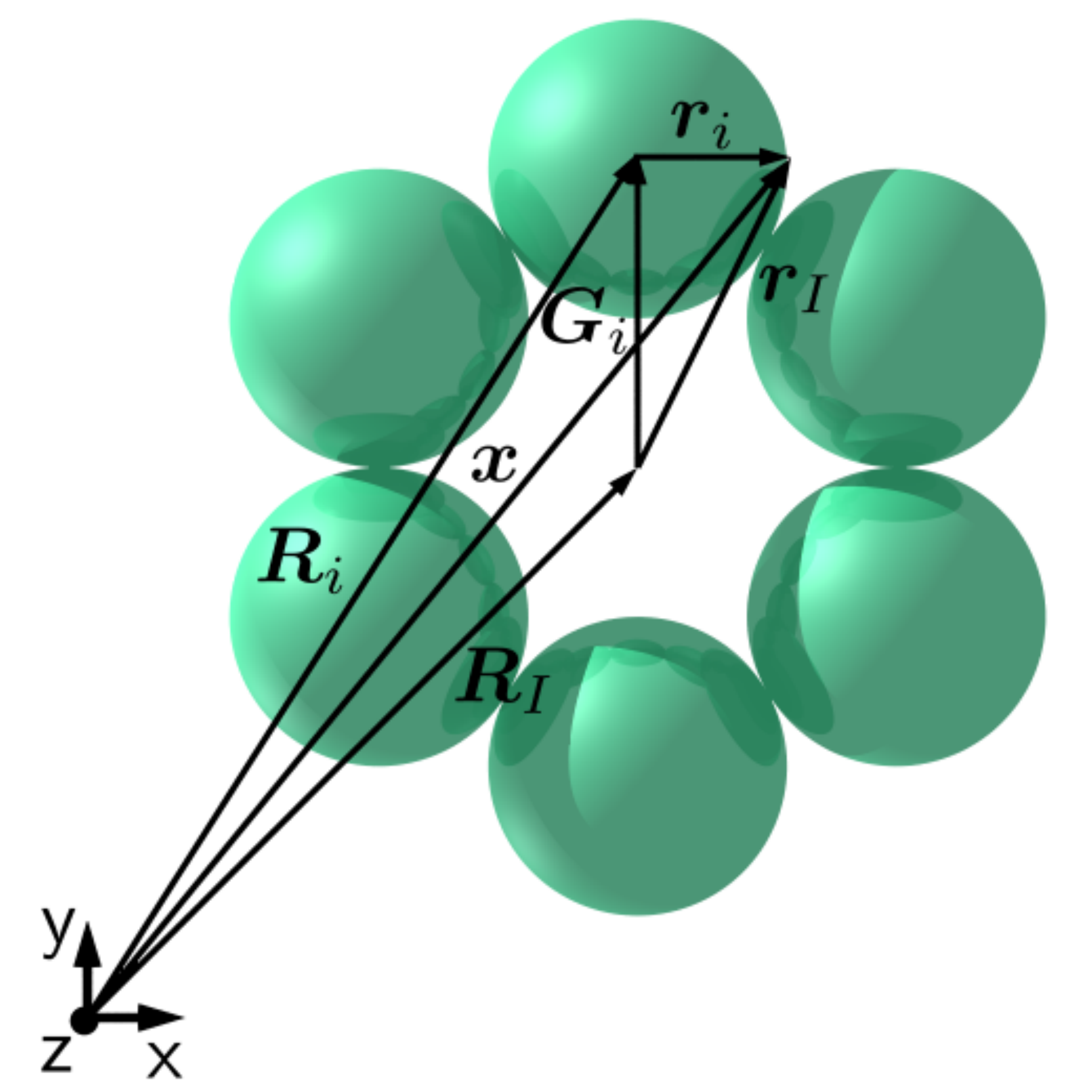}
  \caption{\label{f:beads}Rigid body representation as arbitrary
    collection of spherical beads.}
\end{figure}

For computational simplicity, we consider each particle ($I$) as being
composed of a rigid collection of $n_I$ spherical beads (see
fig.\ref{f:beads}), with position, velocities, and angular velocities
given by $\vec{R}_i$, $\vec{V}_i$, and $\vec{\Omega}_i$. We use upper
and lowercase variables to differentiate between rigid body particles
and the spherical beads used to construct them, as well as the
shorthand $i\in I$ to refer to the $n_I$ beads belonging to the rigid
body $I$. The rigidity constraint on the bead velocities is given by
\begin{align}
  \vec{V}_i &= \vec{V}_I + \vec{\Omega}_I\times\vec{G}_i \label{e:vbead}\\
  \vec{\Omega_i} &= \vec{\Omega}_I\label{e:wbead}
\end{align}
where $\vec{G}_i=\vec{R}_i - \vec{R}_I$ is the distance vector from
the com of the rigid body ($\vec{R}_I$) to the bead's com
($\vec{R}_i$). The rigidity constraint on the position of the beads
requires that the relative distances between any two of them remain
constant. Thus, the $\vec{G}_i$ vectors, expressed within the
reference frame of the particle $\vbody{G}_i$, are constants of motion
\begin{align}
  \vbody{G}_i = \tensor{Q}_I^{\text{t}}\cdot\vec{G}_i =
  \text{constant} \label{e:rbead}
\end{align}
where $\tensor{A}^{t}$ is the matrix transpose of $\tensor{A}$. The
individual positions of the beads can be directly obtained from the
position and orientation of the rigid body to which they belongs
through
\begin{align}
  \vec{R}_i &= \vec{R}_I + \vec{G}_i \label{e:rn_bead} \\
  \vec{G}_i &= \tensor{Q}_I\cdot\vbody{G}_i
\end{align}
The beads should only be considered as a computational bookkeeping
device, used to map the rigid particles onto the computational grid
used to solve the fluid equations of motion. We are free to chose any
representation of the rigid body.

The advantage of this spherical-bead representation is the ease with
which the smooth profile function $\phi_I$ of an arbitrary rigid body
can be defined. We start with the profile function for spherical particles
introduced in ref.~\cite{Nakayama2005a}
\begin{align}
  \phi_i\left(\vec{x}\right) &= \frac{h\left[\left(a + \xi/2\right) - r_i \right]}{
      h\left[\left(a+\xi/2\right) - r_i\right] + h\left[r_i - \left(a
          - \xi/2\right)\right]}\\
    h(x) &= \begin{cases}
      \exp{\left(-\Delta^2 / x^2\right)} & x \ge 0 \\
      0 & x < 0
    \end{cases}
\end{align}
where $\vec{r}_i = \vec{x} - \vec{R}_i$ is the distance vector from
the sphere center to the field point of interest, $a$ is the radius of
the spheres, and $\xi$ is the thickness of the fluid-particle
interface. We then define the smooth profile function of the rigid
body as
\begin{align}
  \phi_I(\vec{x}) &= \sum_{i\in I}\frac{\phi_i(\bm{x})}{\text{max}\left(\sum_{j\in
        I}\phi_j\left(\bm{x}\right), 1\right)}\label{e:phi_bead}
\end{align}
The normalization factor in eq.~\eqref{e:phi_bead} is required to
avoid double-counting in the case of overlap between beads belonging
to the same rigid particle (beads belonging to different particles are
prevented from overlapping by the core potential in $\vec{F}^C$). We
note that this representation of particles as rigid assemblies of
spheres does not impose any constraints on the particle geometries,
because the constituent beads are free to overlap with each other.
\section{Stokes Drag}
If the dispersion under consideration is such that the inertial forces
are negligible compared to the viscous forces, i.e. when the particle
Reynolds number is vanishingly small $\text{Re} = \rho U L / \eta \ll
1$ ($U$ and $L$ being the characteristic velocity and length scales),
the incompressible Navier-Stokes equation (eq.~\eqref{e:ns_host})
reduces to the Stokes equation\cite{Batchelor:2000uu}
\begin{align}
  \eta\nabla^2\vec{u} &= \nabla p + \vec{f}^{\text{Ext}}\label{e:stokes}
\end{align}
with $\vec{f}^{\text{ext}}$ any external forces on the fluid. Due to
the linear nature of this equation, the force (torque) exerted by the
fluid on the particles is also linear in the their
velocities\cite{Kim:2005uv,Dhont:1996ub}
\begin{align}
\tensor{F} &= \tensor{Z}\cdot\tensor{U} \label{e:fzu}\\
\tensor{U} &= \tensor{M}\cdot\tensor{F} \label{e:umf}
\end{align}
where $\tensor{F}=\left(\vec{F}_1,\ldots,\vec{F}_N,\vec{N}_1,\ldots,\vec{N}_N\right)$ and
$\tensor{U}~=~\left(\vec{V}_1,\ldots,\vec{V}_N,\vec{\Omega}_1,\ldots,\vec{\Omega}_N\right)$
are $6N$ dimensional force and velocity vectors, and $\tensor{Z}$ and
$\tensor{M}=\tensor{Z}^{-1}$ are the friction and mobility matrices,
respectively
\begin{align}
  \tensor{Z} &= 
  \begin{pmatrix}
    \tensor{Z}^{tt} & \tensor{Z}^{tr} \\
    \tensor{Z}^{rt} & \tensor{Z}^{rr}
  \end{pmatrix}\\
  &=
  \begin{pmatrix}
    \tensor{\zeta}^{tt}_{11} & \cdots & \tensor{\zeta}^{tt}_{1N} & \tensor{\zeta}^{tr}_{11} &
    \cdots & \tensor{\zeta}^{tr}_{1N} \\
    \vdots & \ddots & \vdots & \vdots & \ddots & \vdots\\
    \tensor{\zeta}^{tt}_{N1} & \cdots & \tensor{\zeta}^{tt}_{NN} & \tensor{\zeta}^{tr}_{N1} & 
    \cdots & \tensor{\zeta}^{tr}_{NN} \\[0.6em]
    \tensor{\zeta}^{rt}_{11} & \cdots & \tensor{\zeta}^{rt}_{1N} & \tensor{\zeta}^{rr}_{11} &
    \cdots & \tensor{\zeta}^{rr}_{1N} \\
    \vdots & \ddots & \vdots & \vdots & \ddots & \vdots \\
    \tensor{\zeta}^{rt}_{N1} & \cdots & \tensor{\zeta}^{rt}_{NN} & \tensor{\zeta}^{rr}_{1N} & 
    \cdots & \tensor{\zeta}^{rr}_{NN}
  \end{pmatrix}
\end{align}
\begin{align}
  \tensor{M} &=
  \begin{pmatrix}
    \tensor{M}^{tt} & \tensor{M}^{tr} \\
    \tensor{M}^{rt} & \tensor{M}^{rr}
  \end{pmatrix}\\
  &=
  \begin{pmatrix}
    \tensor{\mu}^{tt}_{11} & \cdots & \tensor{\mu}^{tt}_{1N} & \tensor{\mu}^{tr}_{11} &
    \cdots & \tensor{\mu}^{tr}_{1N} \\
    \vdots & \ddots & \vdots & \vdots & \ddots & \vdots\\
    \tensor{\mu}^{tt}_{N1} & \cdots & \tensor{\mu}^{tt}_{NN} & \tensor{\mu}^{tr}_{N1} & 
    \cdots & \tensor{\mu}^{tr}_{NN} \\[0.6em]
    \tensor{\mu}^{rt}_{11} & \cdots & \tensor{\mu}^{rt}_{1N} & \tensor{\mu}^{rr}_{11} &
    \cdots & \tensor{\mu}^{rr}_{1N} \\
    \vdots & \ddots & \vdots & \vdots & \ddots & \vdots \\
    \tensor{\mu}^{rt}_{N1} & \cdots & \tensor{\mu}^{rt}_{NN} & \tensor{\mu}^{rr}_{1N} & 
    \cdots & \tensor{\mu}^{rr}_{NN}
  \end{pmatrix}
\end{align}
where the off-diagonal matrices are related through the Lorentz
reciprocal relations by\cite{Kim:2005uv}
\begin{align}
  \tensor{Z}^{rt} &= \left(\tensor{Z}^{tr}\right)^{\text{t}} \\
  \tensor{M}^{rt} &= \left(\tensor{M}^{tr}\right)^{\text{t}}
\end{align}
The symmetric block matrices $\tensor{Z}$ ($\tensor{M}$) are composed
of $3\times 3$ friction (mobility) matrices
$\tensor{\zeta}^{tt}_{ij}$, $\tensor{\zeta}^{tr}_{ij}$,
$\tensor{\zeta}^{rt}_{ij}$, and $\tensor{\zeta}^{rr}_{ij}$ which
couple the translational and rotational motion of particle $i$ with
that of particle $j$. Thus, the whole problem of solving the dynamical
equations of motion for a suspension of spheres in the Stokes regime
reduces to calculating the mobility or friction matrix. For a single
spherical particle, translating (rotating) in an unbounded fluid under
stick-boundary conditions, the translational $\zeta^t$ and rotational
$\zeta^r$ friction coefficients are obtained from an exact solution of
the Stokes equation as\cite{Dhont:1996ub}
\begin{align}
  \zeta^t &= \frac{1}{\mu^t} = 6\pi \eta a \\
  \zeta^r &= \frac{1}{\mu^r} = 8\pi \eta a^3
\end{align}

Exact solutions to the friction or resistance problem of two or three
spherical particles are
known\cite{Jeffrey:1984tu,Kim:1985uq,Kim:1987gl}, but for arbitrary
many-particle systems, the complex non-linear nature of the
hydrodynamic interactions makes it impossible to find a general
solution. However, several methods have been developed to obtain
accurate estimates for the mobility and friction matrices. Two of the
most popular approaches are the method of reflections and the method
of induced forces\cite{Dhont:1996ub,Cichocki:1994gl}. The former
relies on a power series expansion of the flow field, in terms of the
inverse particle distances ($a/r_{ij}$), while the latter uses a
multipole expansion of the force densities induced at the particle
surface, with the truncation scheme determined by the angular
dependence of the flow field ($l=1,\ldots, L$). As an example, the
popular Rotne-Prager (RPY) approximation to the mobility tensor can be
obtained using the method of reflections, by truncating the
hydrodynamic interactions to third order, which corresponds to a
pair-wise representation
\begin{align}
  \tensor{\mu}_{ij}^{tt} &= \begin{cases}
    \mu^t \tensor{I} & i=j \\
    \mu^t \left[\frac{3}{4}\left(\tensor{I} +
        \uvec{r}_{ij}\uvec{r}_{ij}\right) 
      +  \frac{1}{2}\left(\frac{a}{r_{ij}}\right)^3\left(
        \tensor{I} - 3\uvec{r}_{ij}\uvec{r}_{ij}
      \right)\right] & i\ne j
  \end{cases}
\end{align}
We will compare our DNS results with those obtained using the method
of induced forces, using the freely available \texttt{HYDROLIB}
library\cite{Hinsen:1995ty}. Calculations of the friction
coefficients for a sedimenting array of spherical clusters, using the
induced force method truncated to third order ($L=3$), gives an error
of less than $1\%$ with respect to the experimental
results\cite{Cichocki:1995bsa}. We therefore consider these values as
the exact solution to the Stokes equation (SE). Finally, we note that
neither the method of reflections nor the induced force method is able
to directly take into account lubrication forces, caused by the
relative motion of particles at short distances, as they require the
high-order terms to be included in both expansions. Following
Durlofsky et al.\cite{Durlofsky:1987vh}, these contributions are
usually added to the friction matrix by assuming a pairwise
superposition approximation. In this work we will consider only the
collective motion of a rigid agglomerate of spheres, so that
lubrication effects need not be considered.

\section{Results}
In what follows we report the friction coefficients for a variety of
non-spherical particles under steady translation (rotation) through a
fluid. The numerical simulations are performed in three dimensions
under periodic boundary conditions. The modified Navier-Stokes
equation (eq.\eqref{e:ns}) is discretized with a dealiased Fourier
spectral scheme in space and an Euler scheme in time. The motion of
the particles is integrated using a second order Adams-Bashforth
scheme. The lattice spacing $\Delta$ is taken as unit of length, the
unit of time is given by $\rho \Delta^2 / \eta$, with $\rho=1$ and
$\eta=1$ the density and viscosity of the fluid. The integration time
step is $h=7.5 \times 10^{-2}$. We only consider neutrally bouyant
particles, $\rho_p = \rho$, so gravity effects are not considered
Furthermore, we are only interested in single particle motion, so the
particle-particle interactions and external field contributions can be
ignored $\vec{F}^{\text{Ext}}=\vec{F}^{\text{C}} = 0$.

The rigid particles are constructed as rigid agglomerates of
non-overlapping spherical beads of equal radius $a$. We use a bead
radius of $a = 2, 4$ and a system size of $L = 128, 256$ (depending on
the particle geometry); the interfacial thickness is $\zeta = 2$ in
all cases. The particle velocity $\stensor{U}^{\alpha}$ is fixed
throughout the simulation and the steady-state forces
$\stensor{F}^{\beta}$, which are purely hydrodynamic in nature, are
measured in order to obtain the friction coefficients
$\zeta^{\alpha\beta}$. All results are given in terms of the kinetic
form factors $\stensor{K}^{\alpha\beta}$, defined as
\begin{align}
  \tensor{Z} = \begin{pmatrix}
    \tensor{K}^{tt} \zeta^t\phantom{/\zeta^t} & \tensor{K}^{tr} \zeta^r/\zeta^t \\
    \tensor{K}^{rt} \zeta^r/\zeta^t & \tensor{K}^{rr} \zeta^r\phantom{/\zeta^t}
  \end{pmatrix}
\end{align}
such that $\tensor{K}^{tt}$ ($\tensor{K}^{rr}$) expresses the force
(torque) on the agglomerates relative the force (torque) experienced
by a spherical particle of equal volume (radius $a_e$) moving
(rotating) at the same velocity (angular velocity) and under the same
boundary conditions. We use the general term, friction tensor or
matrix, to refer to both $\tensor{Z}$ and $\tensor{K}$.

\subsection{Spherical Agglomerates}
\begin{figure}[ht!]
  \begin{subfigure}[b]{0.15\textwidth}
    \centering
    \includegraphics[height=0.85\textwidth]{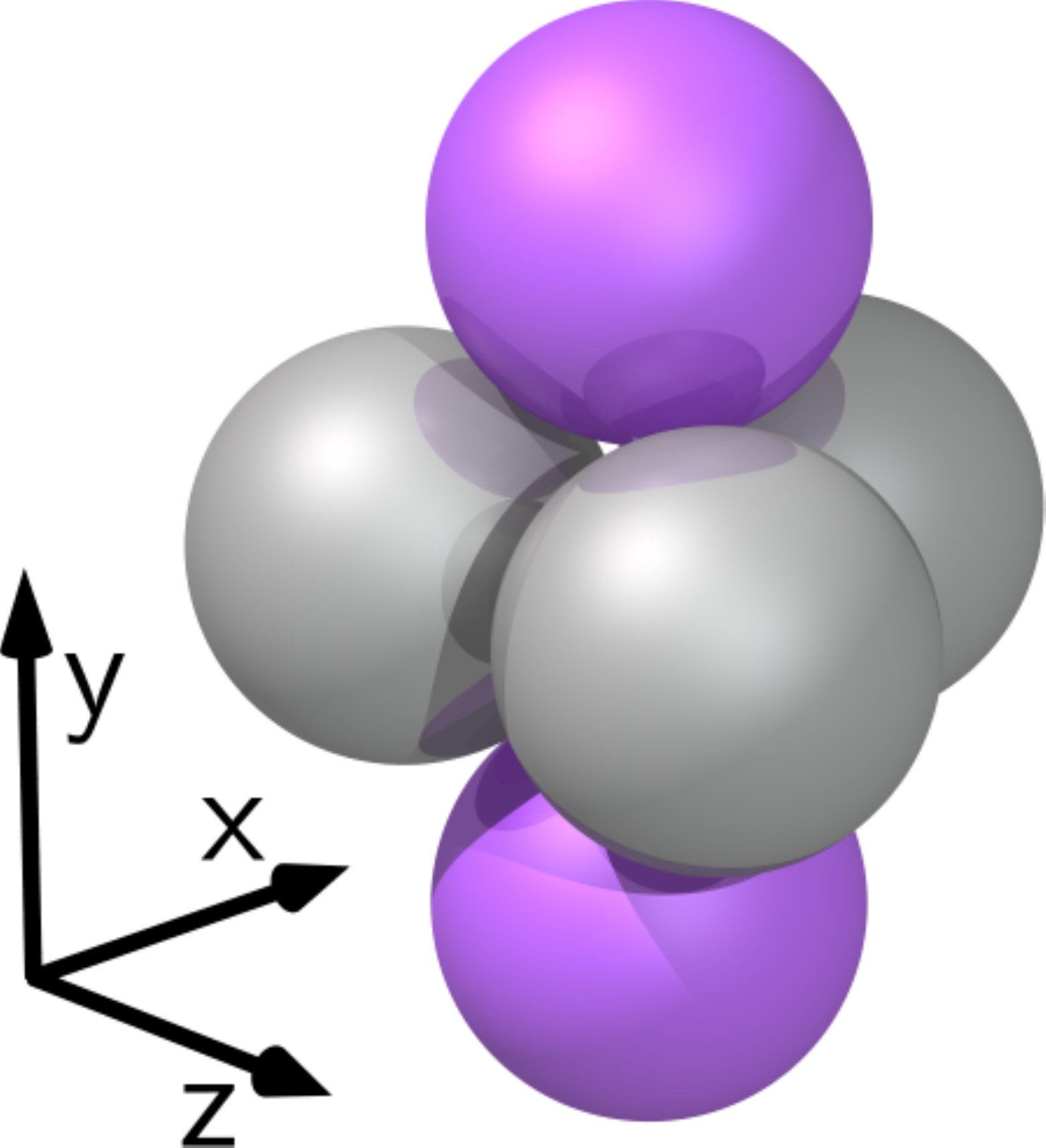}
    \caption{hcp$_5$}
  \end{subfigure}
  \begin{subfigure}[b]{0.15\textwidth}
    \centering
    \includegraphics[height=\textwidth]{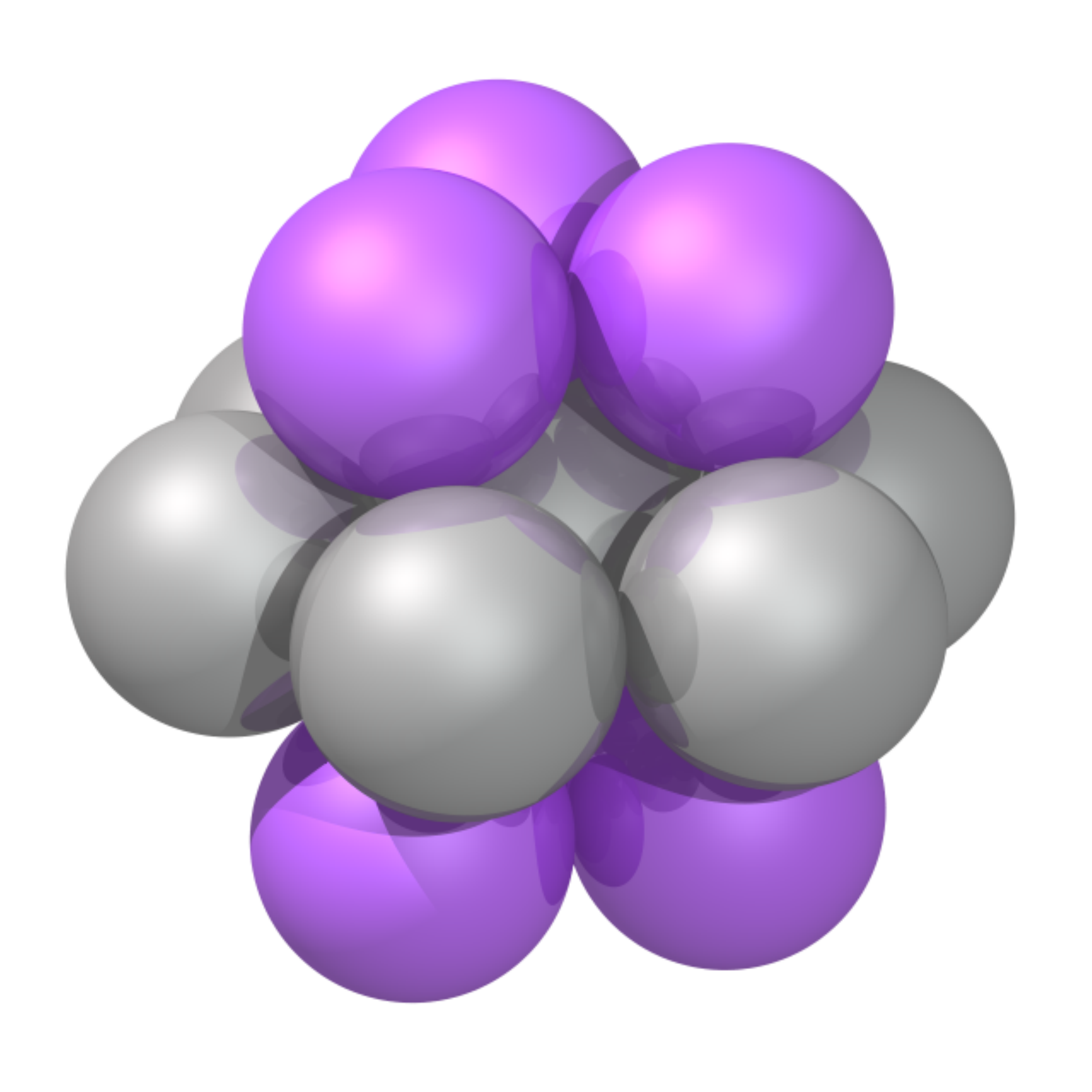}
    \caption{hcp$_{13}$}
  \end{subfigure}
  \begin{subfigure}[b]{0.15\textwidth}
    \centering
    \includegraphics[height=\textwidth]{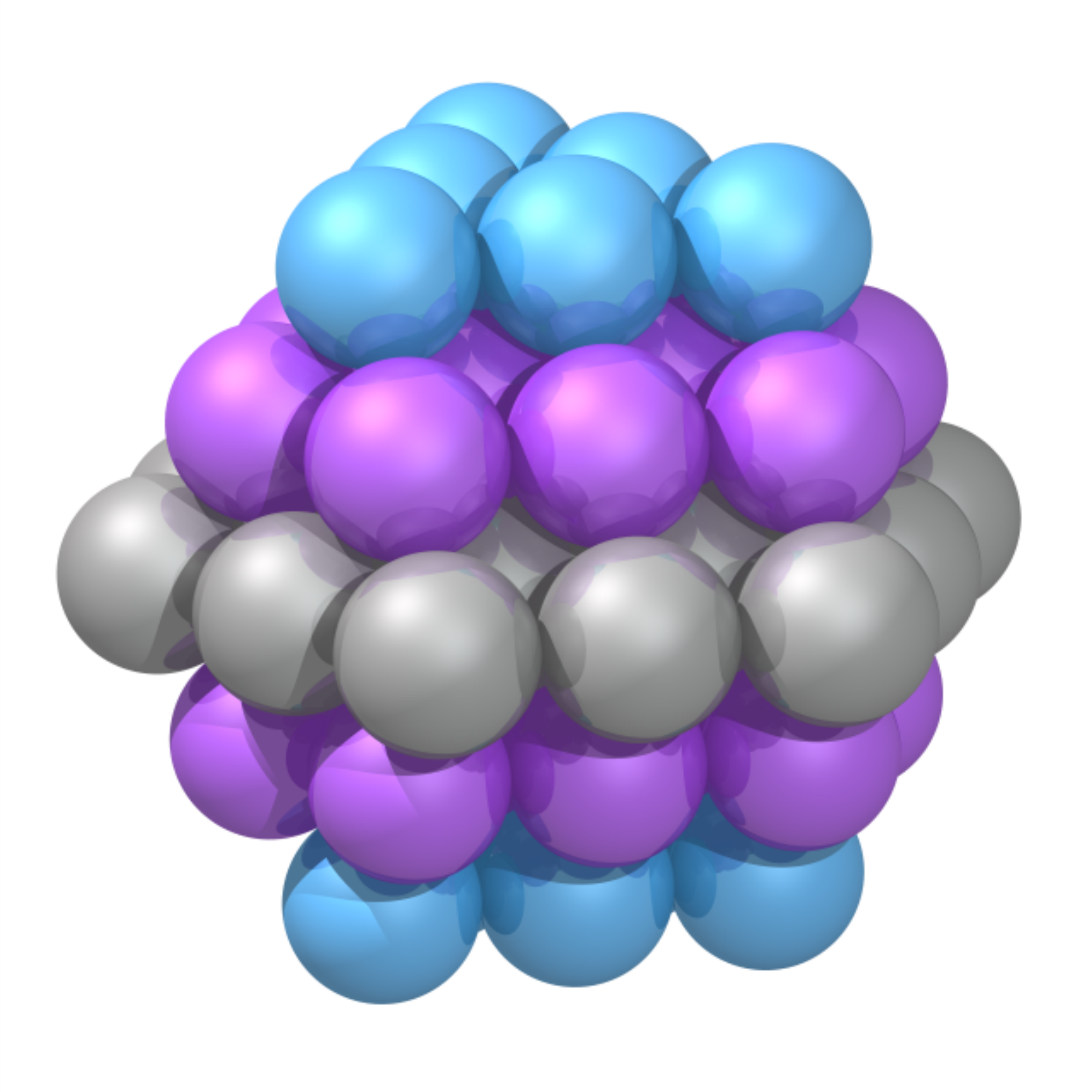}
    \caption{fcp$_{55}$}
  \end{subfigure}
  \begin{subfigure}[b]{0.15\textwidth}
    \centering
    \includegraphics[height=\textwidth]{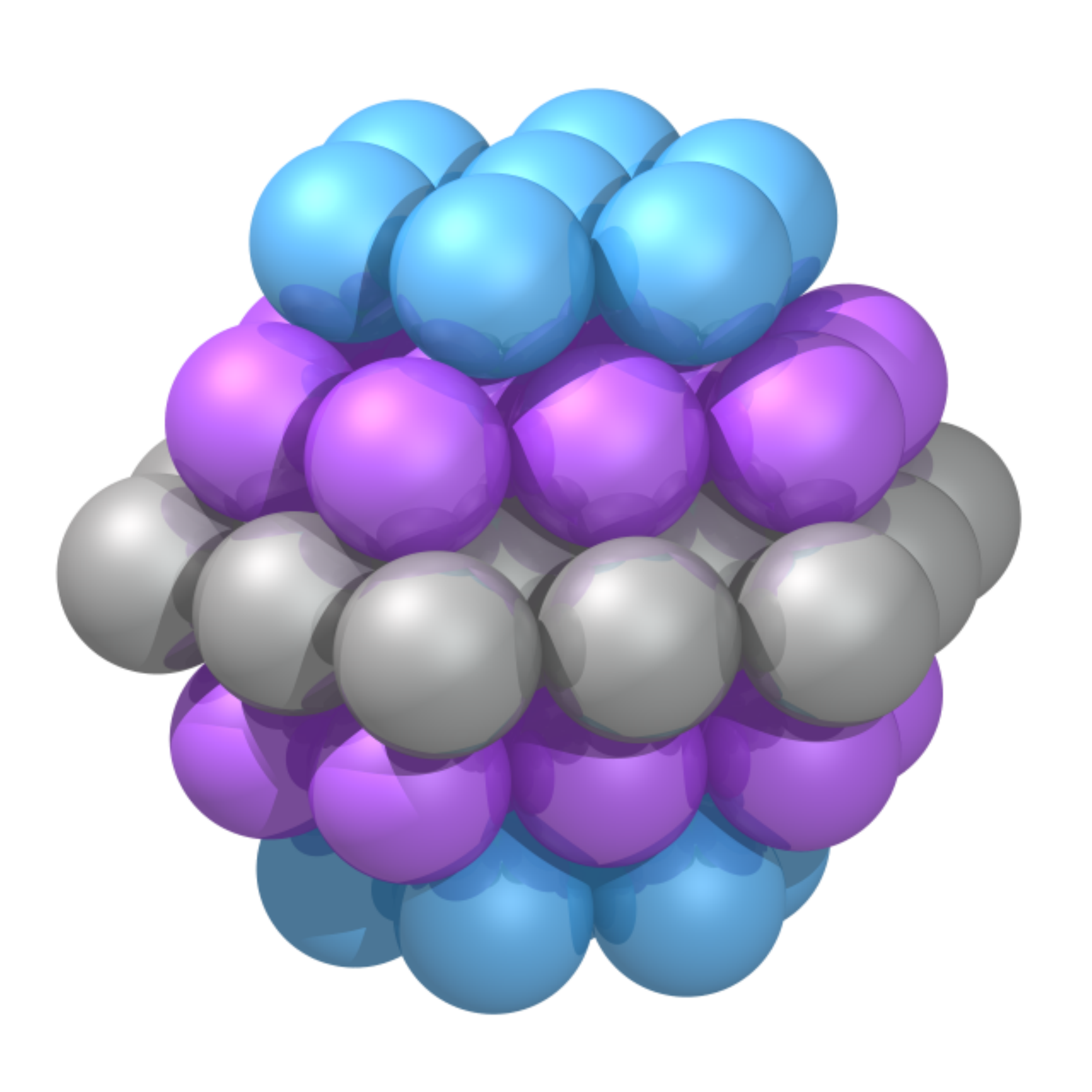}
    \caption{hcp$_{57}$}
  \end{subfigure}
  \begin{subfigure}[b]{0.15\textwidth}
    \centering
    \includegraphics[height=\textwidth]{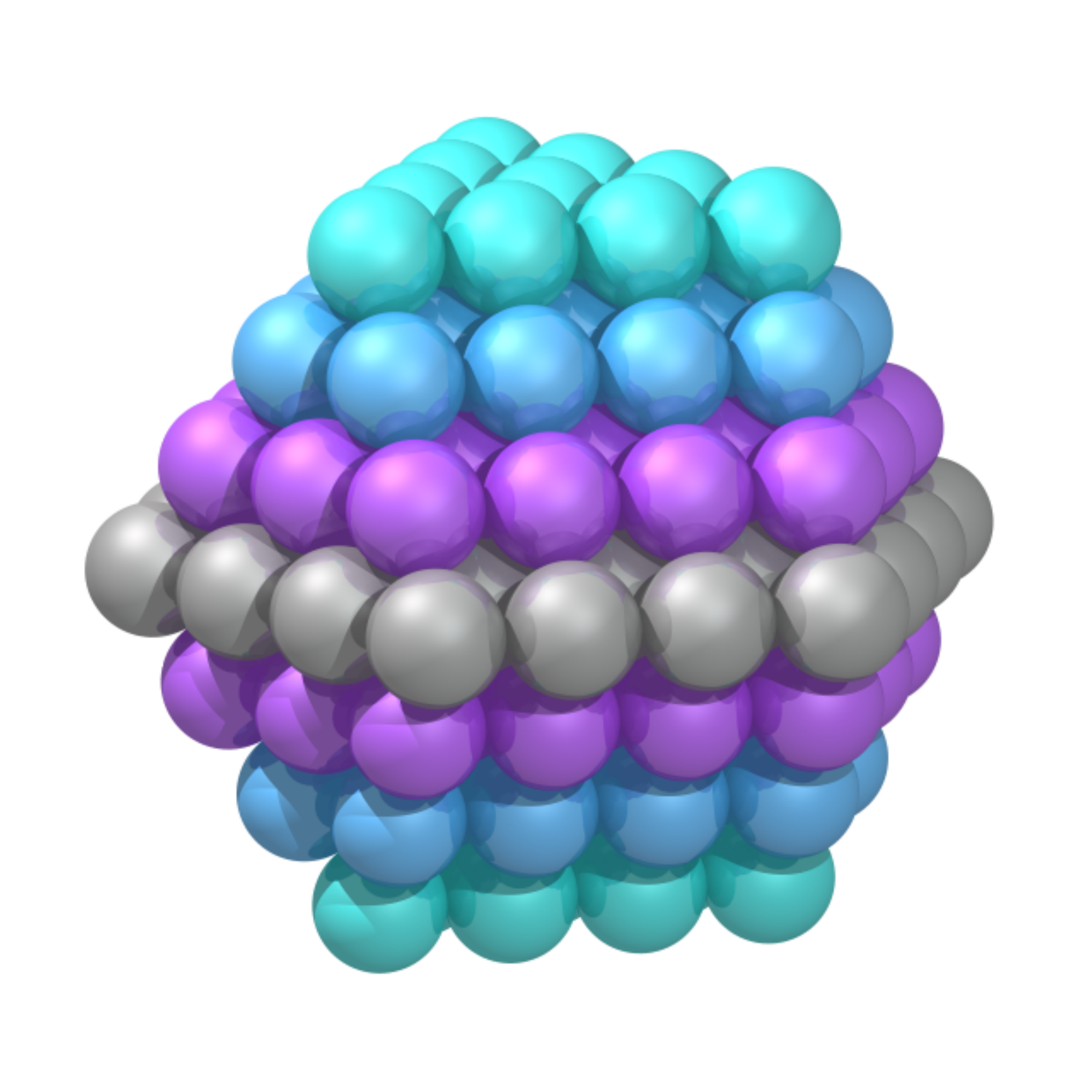}
    \caption{hcp$_{147}$}
  \end{subfigure}
  \begin{subfigure}[b]{0.15\textwidth}
    \centering
    \includegraphics[height=\textwidth]{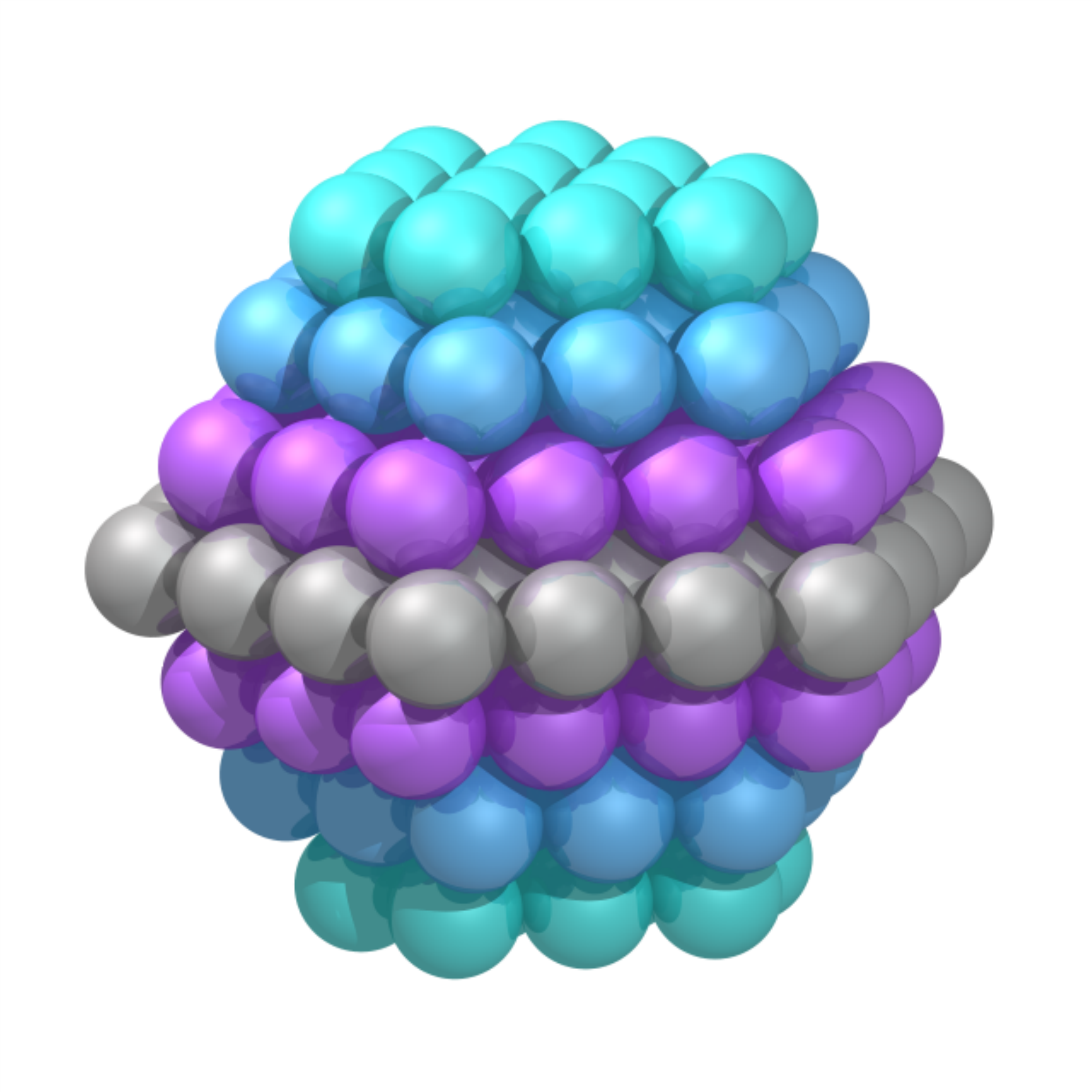}
    \caption{hcp$_{153}$}
  \end{subfigure}
  \begin{subfigure}[b]{0.15\textwidth}
    \centering
    \includegraphics[height=\textwidth]{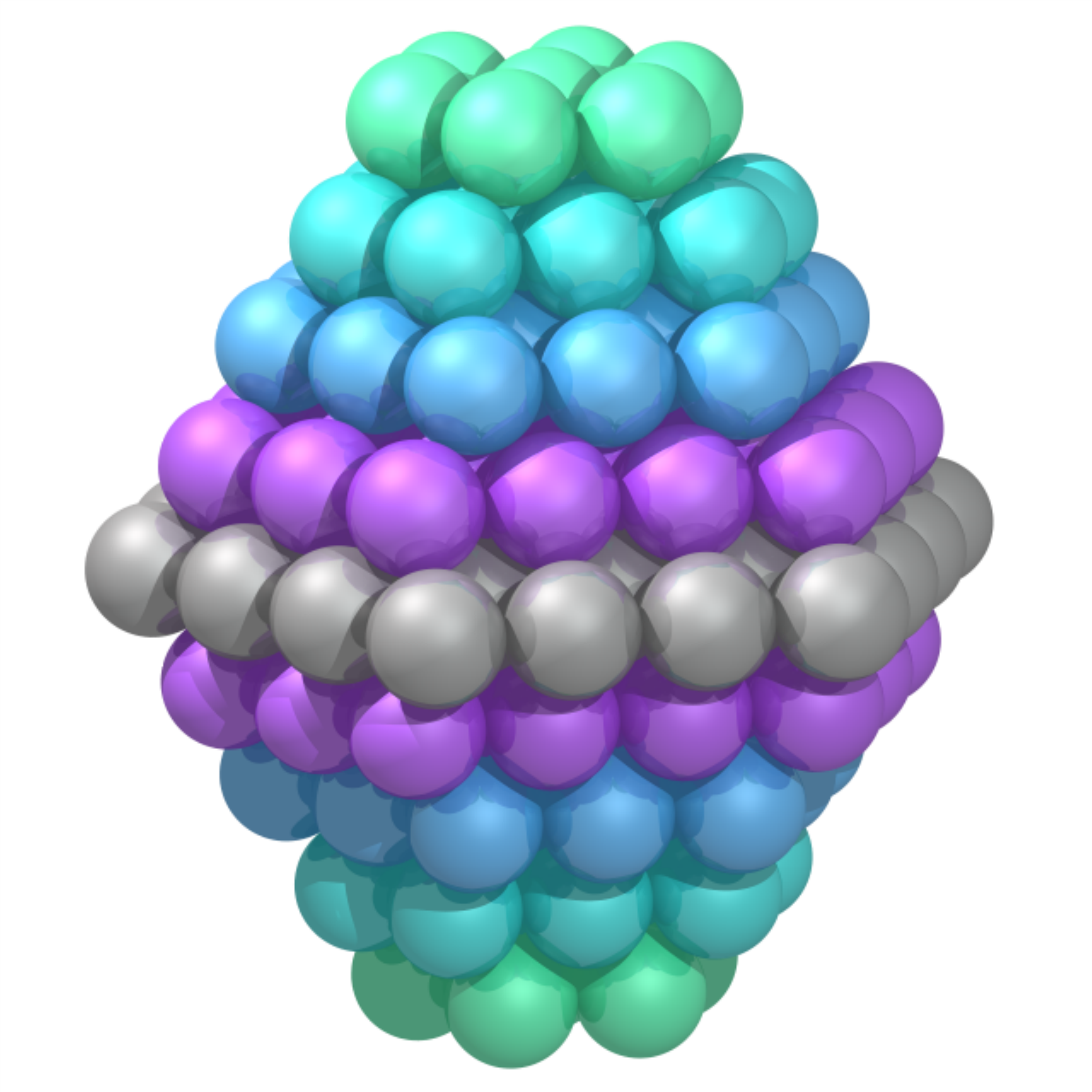}
    \caption{hcp$_{167}$}
  \end{subfigure}
  \caption{\label{f:spherical}(color online) Closed-packed (hcp and
    fcc) arrays of spherical particles. Colors are assigned to the
    individual layers as a visual guide.}
\end{figure}
We begin by measuring the friction coefficients of the spherical
agglomerates studied experimentally in
ref~\cite{Cichocki:1995bsa}. The agglomerates where
constructed to obtain semi-spherical geometries by gluing together from two
to $167$ spherical beads of equal radius within a closest-packing
arrangement (using either hcp or fcc lattices). The spherical
agglomerates for $N\ge 5$ are shown in Fig.~\ref{f:spherical}. The
system parameters are $a=2$, $L=128$, and $V^{\alpha}=0.02$
($\alpha=y,z$), which gives a Reynolds number of $\le 0.44$. Given the
symmetry of the particles, the friction matrix is diagonal, with only
two distinct coefficients
\begin{align}
  \tensor{K}^{tt} = \begin{pmatrix}
    \stensor{K}^{zz} & 0 & 0 \\
    0 & \stensor{K}^{yy} & 0 \\
    0 & 0 & \stensor{K}^{zz}
  \end{pmatrix}
\end{align}
\begin{figure}[ht!]
  \centering
  \includegraphics[width=0.48\textwidth]{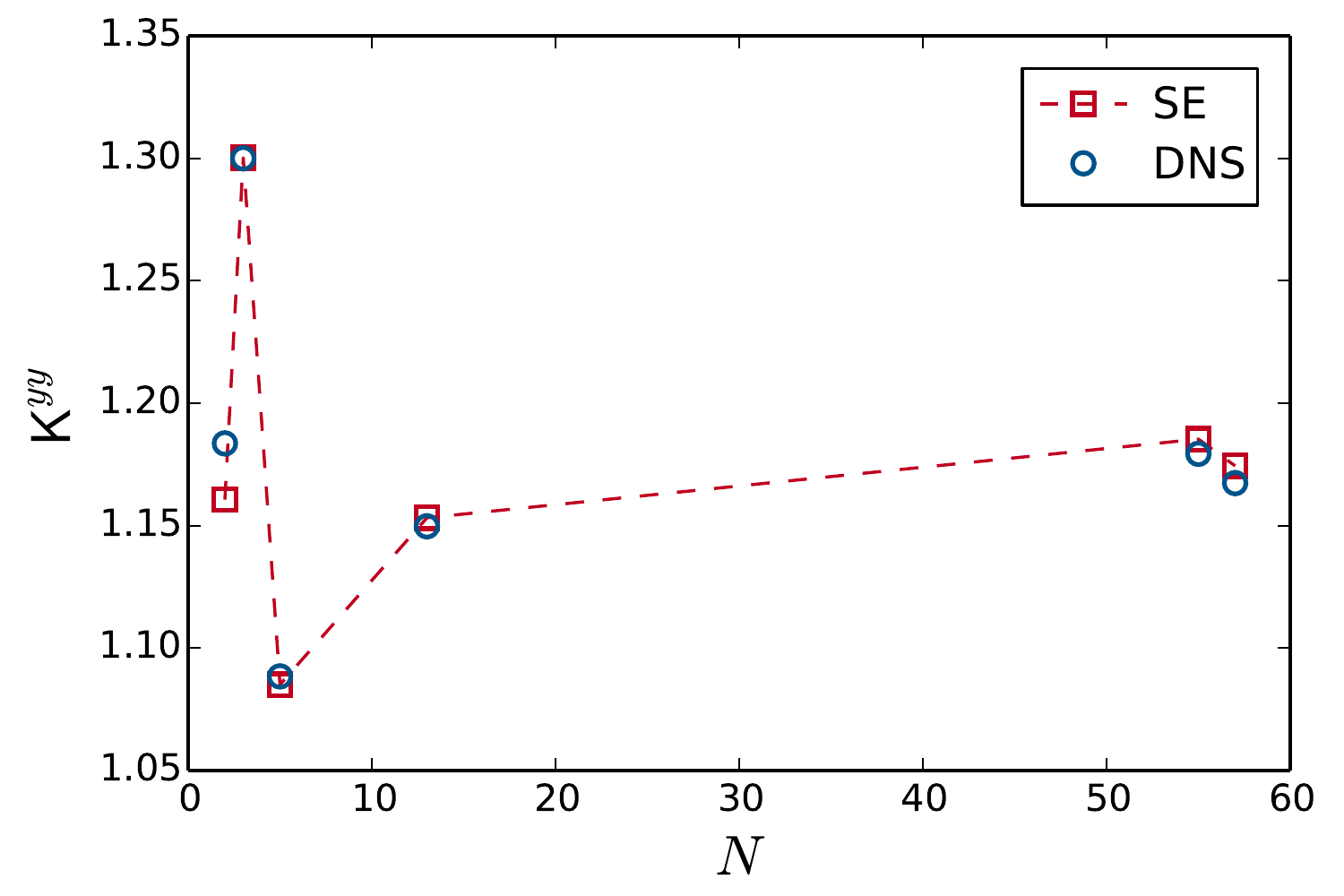}
  \caption{\label{f:spherical_k}(color online) Friction coefficients
    $\stensor{K}^{yy}$ for several closed-packed arrays of spherical
    particles as a function of size (number of spheres). Results
    obtained from DNS calculations (circle) are compared to the exact
    solution to the Stokes equation (square).}
\end{figure}
\begin{table}[ht!]
  \begin{tabular}{l cccccc}
    $N$ & $a/a_e$ & $\stensor{K}_{SE}^{yy}$ &$\stensor{K}_{DNS}^{yy}$ & 
    $\stensor{K}_{SE}^{zz}$ & $\stensor{K}_{DNS}^{zz}$ \\
    \hline
    2   & $1.26$ &$1.161$ & $1.184$ & $1.025$ &$1.051$\\
    3   & $1.44$ &$1.300$ & $1.300$ & $1.074$ &$1.081$\\
    5   & $1.71$ &$1.085$ & $1.088$ & $1.172$ &$1.173$\\
    13  & $2.35$ &$1.153$ & $1.150$ & $1.153$ &$1.148$\\
    55  & $3.80$ &$1.185$ & $1.179$ & $1.185$ &$1.175$\\
    57  & $3.84$ &$1.174$ & $1.167$ & $1.178$ &$1.167$\\
    147 & $5.28$ &---    & $1.198$ & ---   &$1.195$\\
    153 & $5.35$ &---    & $1.184$ & ---    &$1.186$\\
    167 & $5.50$ &---    & $1.156$ & ---    &$1.227$
  \end{tabular}
  \caption{\label{t:spherical} Friction coefficients for motion
    parallel and perpendicular to the vertical $y$-axis for all the
    closed-packed arrays, as given by our DNS method and the exact
    solution to the SE.}
\end{table}

Precise experimental measurements are available for these systems, but
they should not be compared directly to our simulation results due to
the mismatch in boundary conditions, particularly for the larger
agglomerates. The friction coefficients for movement along the
vertical $y$-axis are given in fig~\ref{f:spherical_k}, where they are
compared to the exact (SE) results. The complete set of values for the
two independent friction coefficients are given in
table~\ref{t:spherical}. For the larger systems, $N\ge 147$, the
\texttt{HYDROLIB} library does not converge if periodic boundary
conditions are used, so no reference data is given. Our results show
excellent agreement with the available SE values, differentiating
between nearly identical agglomerates which differ in volume by only a
few percent. In all cases, the difference between our results and the
reference values are less than $2\%$, which is the comparable with the
error estimates of the actual experiments.
\subsection{Non-Spherical Agglomerates}
\begin{figure}[ht!]
  \centering
  \begin{subfigure}[b]{0.2\textwidth}
    \centering
    \includegraphics[height=0.08\textheight]{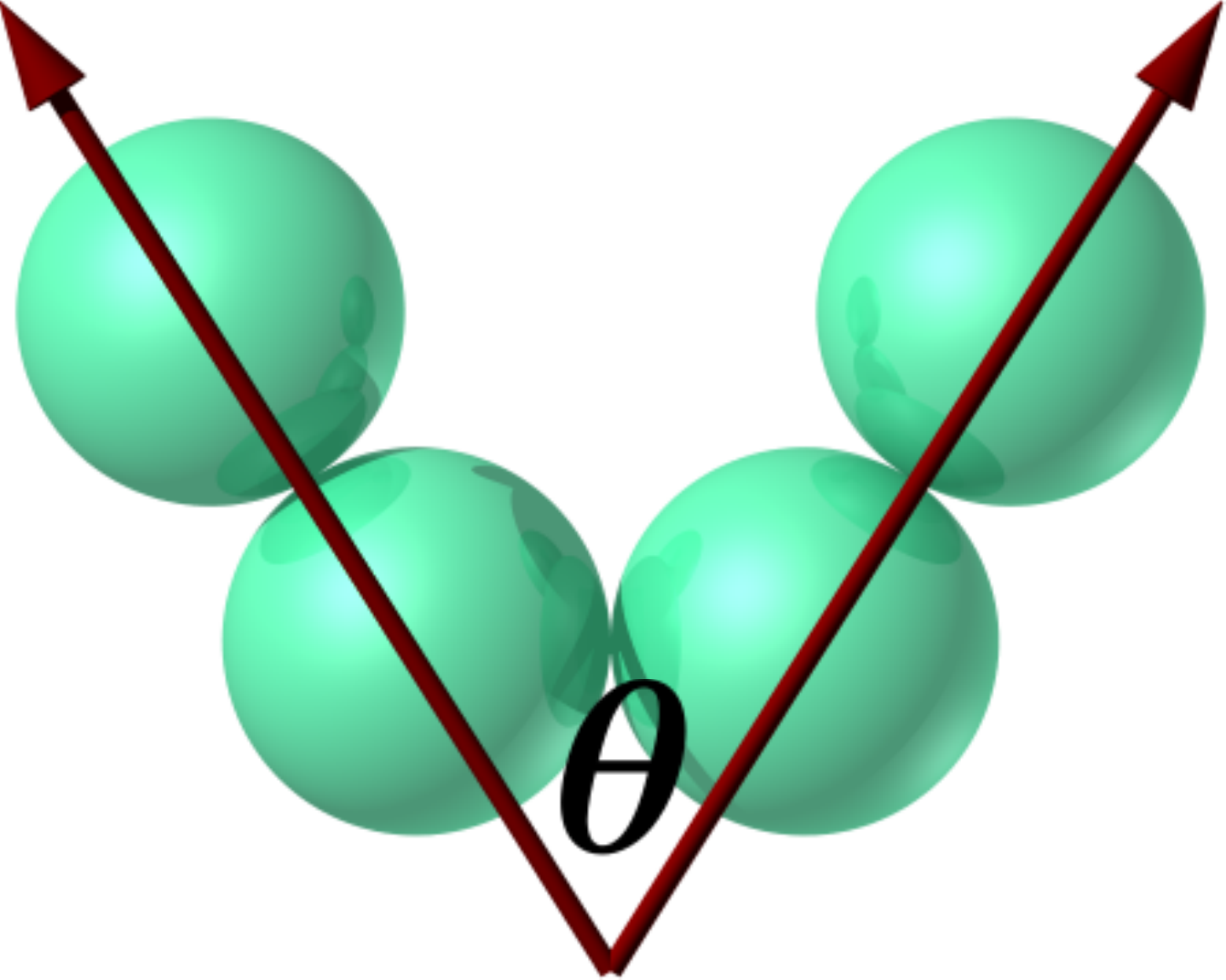}
    \caption{v-shaped (even)}
  \end{subfigure}
  \begin{subfigure}[b]{0.2\textwidth}
    \centering
    \includegraphics[height=0.1\textheight]{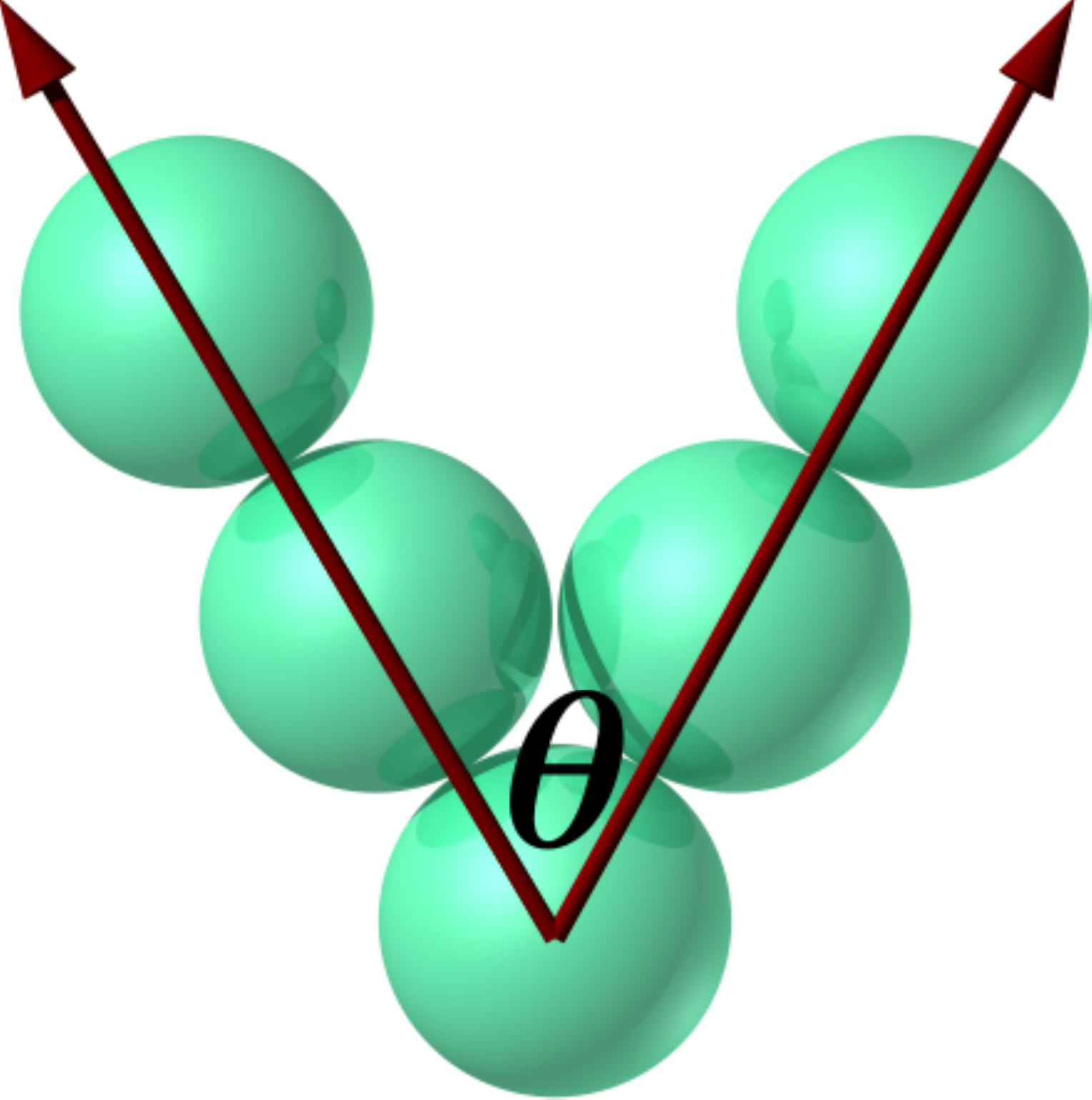}
    \caption{v-shaped (odd)}
  \end{subfigure}
  \begin{subfigure}[b]{0.2\textwidth}
    \centering
    \includegraphics[width=\textwidth]{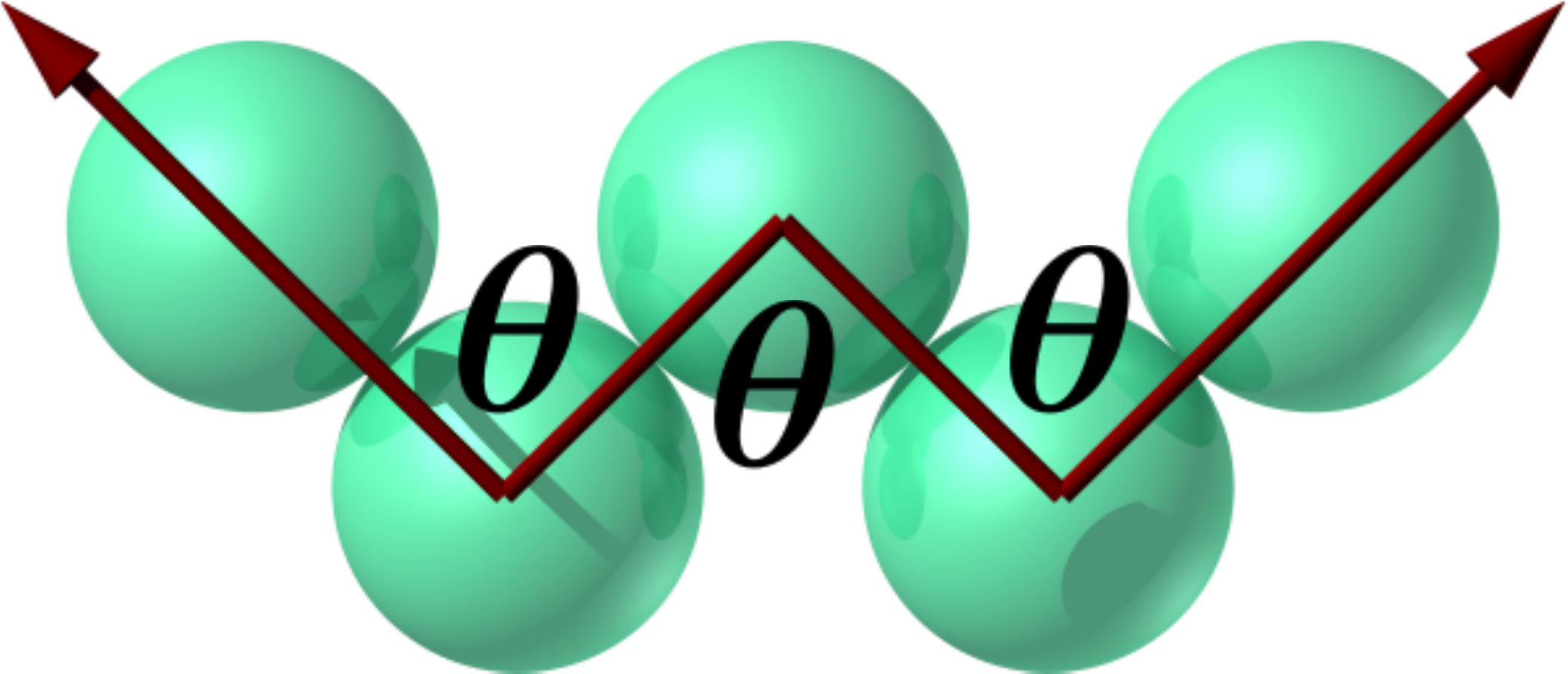}
    \caption{w-shaped}
  \end{subfigure}
  \begin{subfigure}[b]{0.2\textwidth}
    \centering
    \includegraphics[height=0.12\textheight]{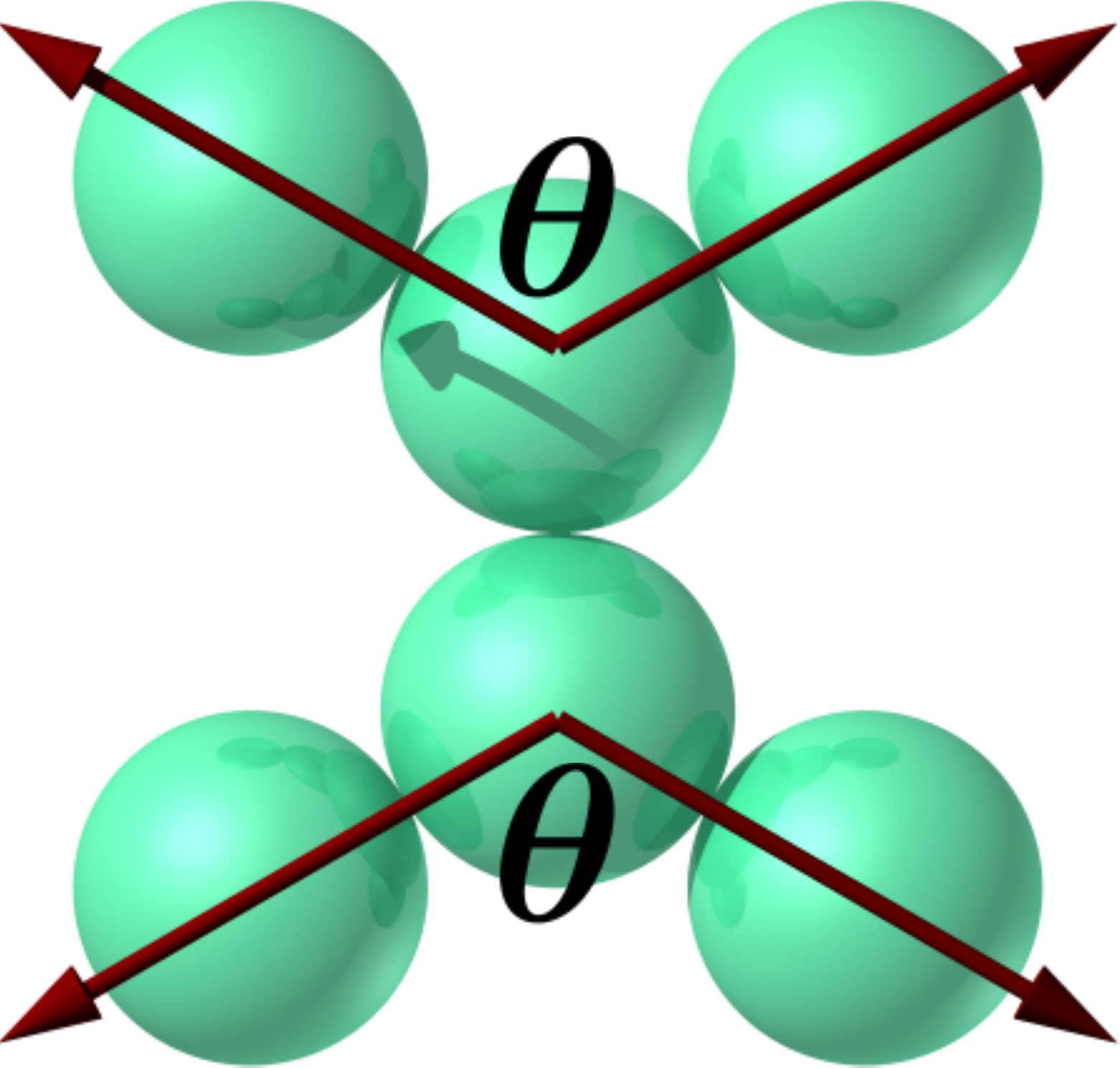}
    \caption{h-shaped}
  \end{subfigure}
  \begin{subfigure}[b]{0.2\textwidth}
    \centering
    \includegraphics[height=0.12\textheight]{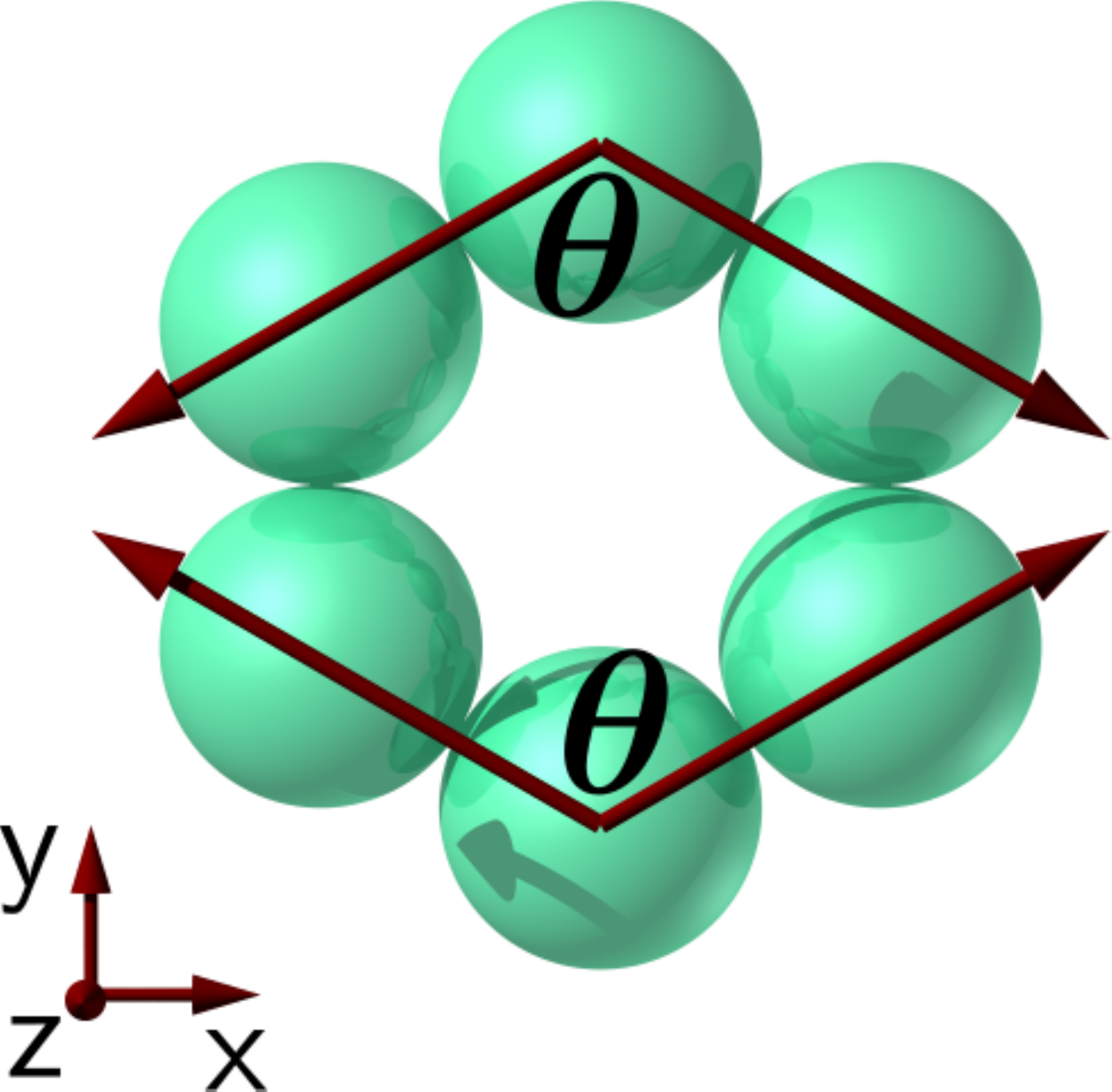}
    \caption{hexagonal}
  \end{subfigure}
  \begin{subfigure}[b]{0.2\textwidth}
    \centering
    \includegraphics[height=0.15\textheight]{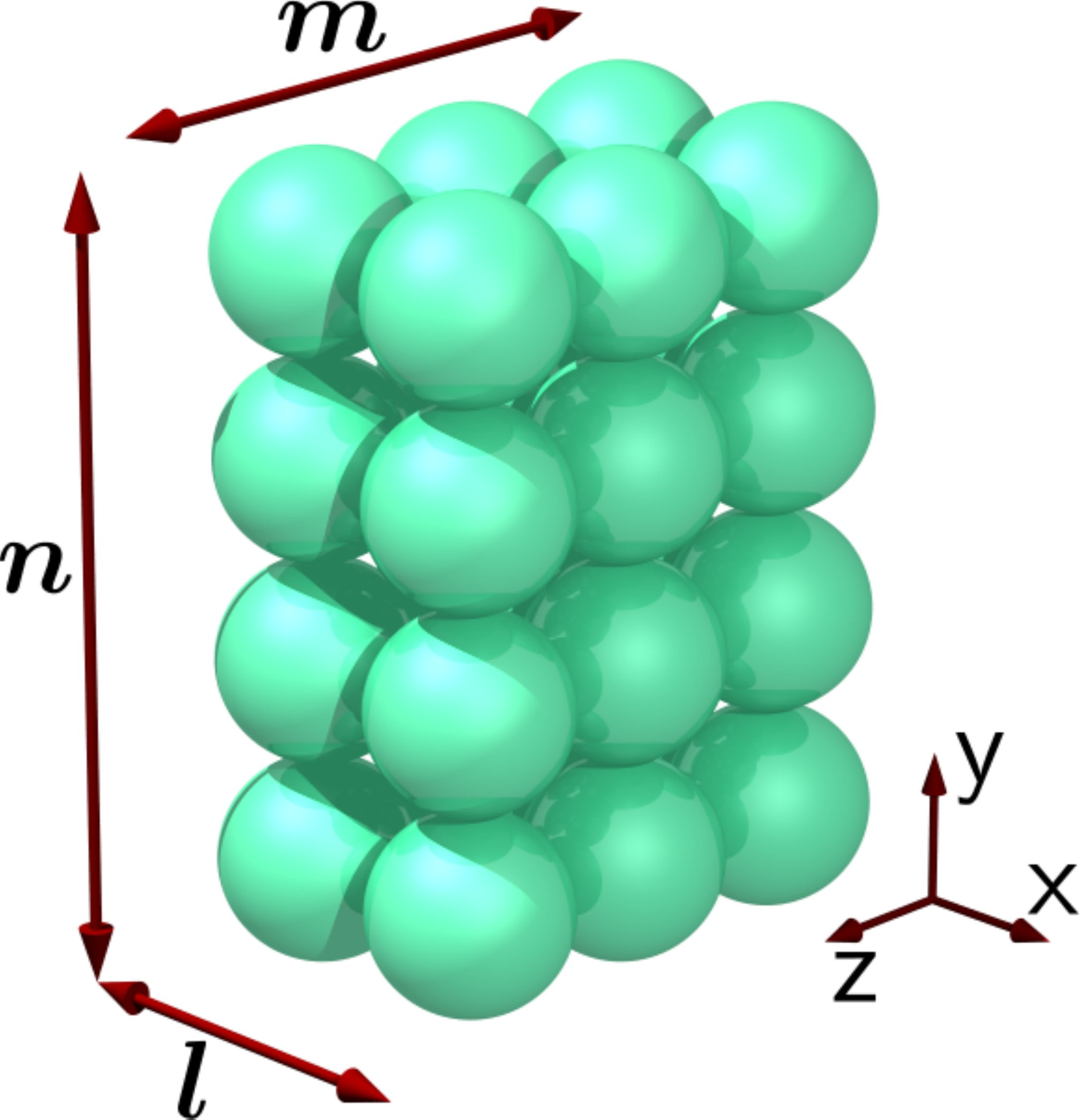}
    \caption{rectangular}
  \end{subfigure}
  \caption{\label{f:nonspherical}(color online) Non-spherical
    agglomerates for various regular shaped geometries.}
\end{figure}

We now consider the friction coefficients for a series of
non-spherical regular-shaped agglomerates. The simulation protocol is
exactly the same as for the spherical agglomerates considered above.
In total, we study six different families of configurations, shown
schematically in fig.~\ref{f:nonspherical}, v-shaped, w-shaped,
h-shaped, hexagonal and rectangular arrays. The v- and w-shaped
configurations vary in the number of particles ($N=3,4,5$), as well as
the branching angle $\theta$. For the h-shaped and hexagonal
configurations only the branching angle is varied, the number of
particles is fixed to six. For the rectangular $(l\times m)\times n$
arrays, the maximum number of particles in any dimension is four. In
total, we have considered $84$ different geometric configurations.
Details on the construction of the agglomerates, as well as
experimental data for the kinematic form factors, can be found in
ref.~\cite{Niida:1997ty}.
The total friction matrix $\tensor{K}$ is block diagonal for all the
geometrical configurations considered here, except for the case of v-
and w-shaped particles, for which a slight coupling between
translation and rotation can be observed: $(\stensor{K}^{rt})^{xz}$
and $(\stensor{K}^{rt})^{zx}\ne 0$. For the moment, however, we
consider only translational motion. Due to the symmetry of the
particles, the friction matrix $\tensor{K}^{tt}$ is again diagonal
\begin{align}
  \tensor{K}^{tt} &=
  \stensor{K}^{\alpha\beta}\tensor{\delta}_{\alpha\beta}
\end{align}

\begin{figure}[ht!]
  \centering
  \begin{subfigure}[b]{0.42\textwidth}
    \includegraphics[width=\textwidth]{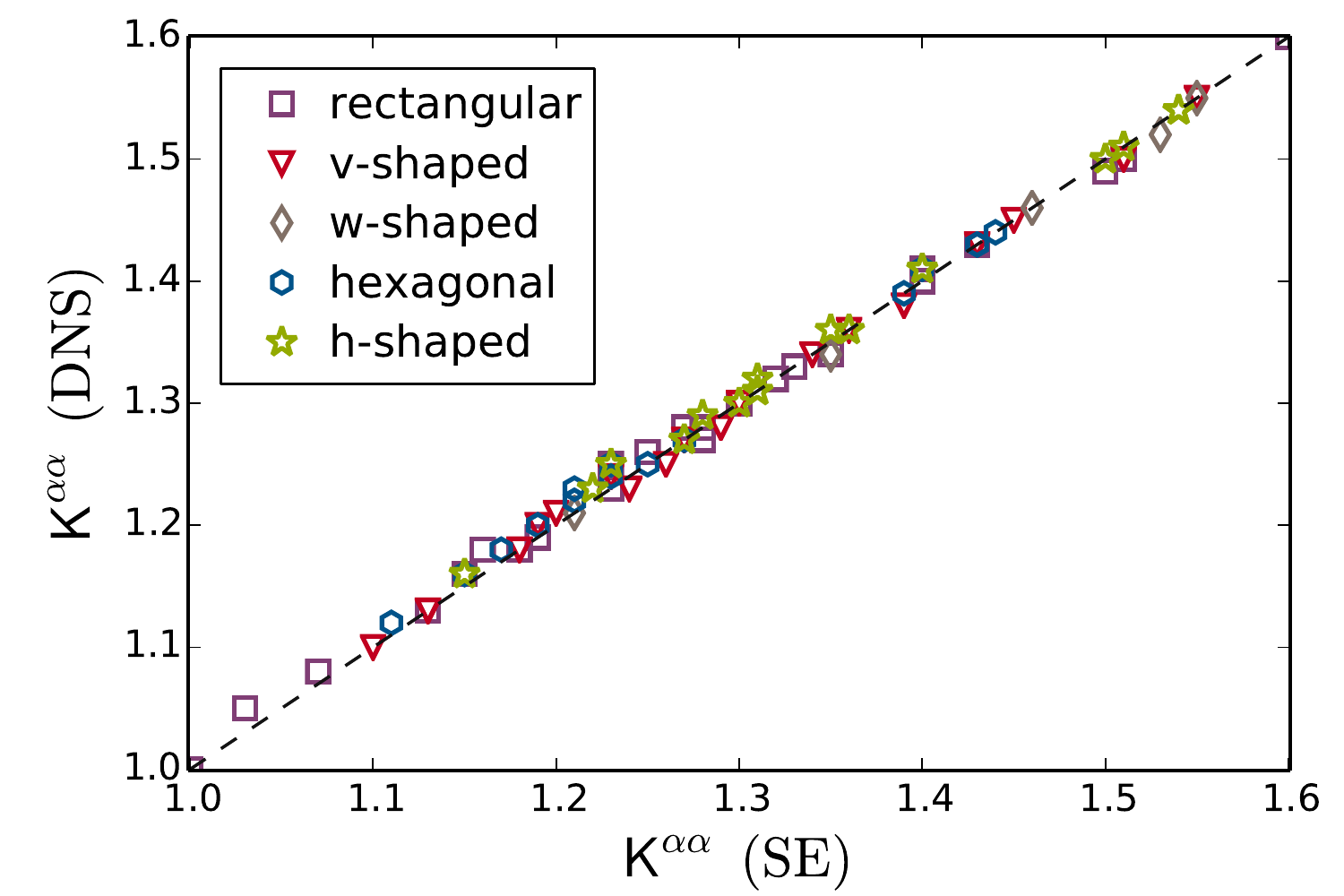}
    \caption{\label{f:nonspherical_k_all} DNS versus SE}
  \end{subfigure}
  \begin{subfigure}[b]{0.42\textwidth}
    \includegraphics[width=\textwidth]{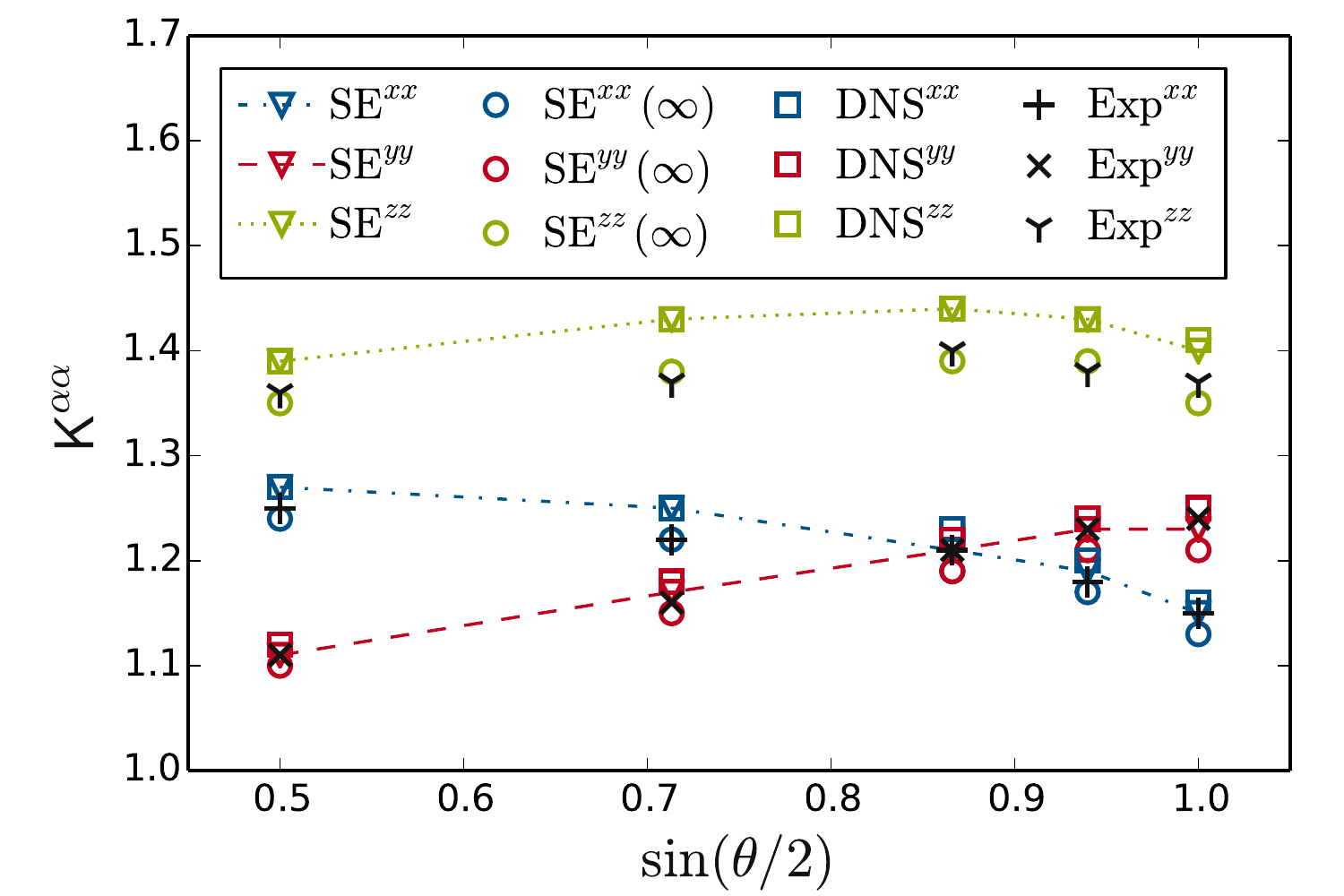}
    \caption{\label{f:nonspherical_k_hex} hexagonal}
  \end{subfigure}
  \begin{subfigure}[b]{0.42\textwidth}
    \includegraphics[width=\textwidth]{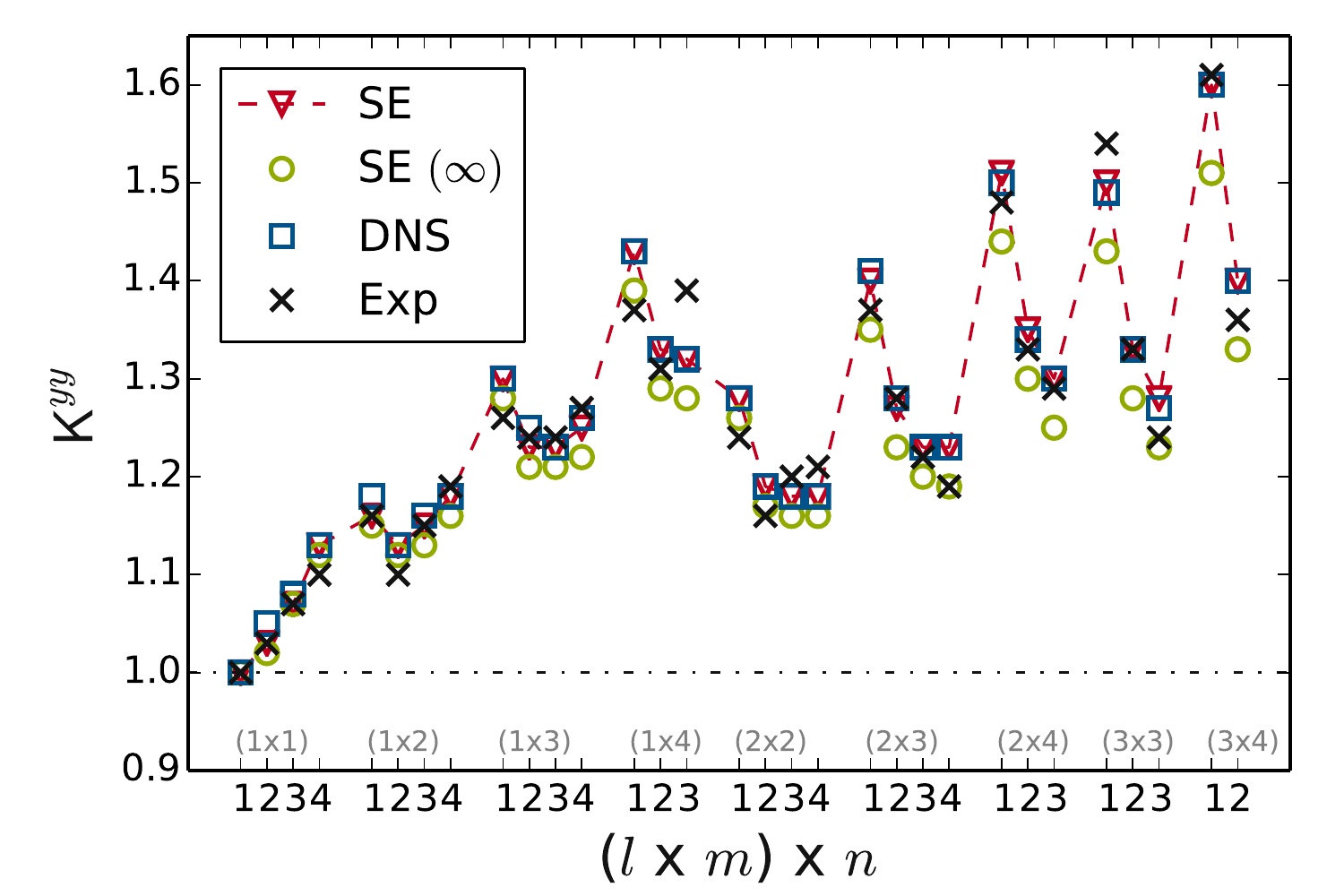}
    \caption{\label{f:nonspherical_k_rect} rectangular}
  \end{subfigure}
  \caption{\label{f:nonspherical_k}(color online) Friction
    coefficients for various non-spherical regular-shaped
    agglomerates. (a) All the friction coefficients computed with the
    DNS method as a function of the exact SE values. (b) The friction
    coefficients for the hexagonal shaped agglomerates as a function
    of their vertical dimension. (c) Vertical friction coefficients
    for rectangular $(l\times m)\times n$ arrays as a function of
    height $n$. Values obtained from the exact solution to the SE, for
    both periodic and unbounded $(\infty)$ systems, as well as
    experimental values, are also shown in (b) and (c).}
\end{figure}

We have computed the friction coefficients for motion parallel to the
vertical $y$ axis for all the systems, the full friction matrix is
only measured for the h-shaped and hexagonal agglomerates. Our results
are summarized in fig.~\ref{f:nonspherical_k}, along with the
experimental data, and the exact solutions for both a periodic and an
infinite system. Although our DNS results should only be compared with
the SE solutions under equivalent boundary conditions, the periodicity
effects for the systems considered here are small, being of the same
order of magnitude as the errors in the experiments. The relative
error of the DNS results (compared to experiments) is less than $5\%$
for all configurations. A comparison of our results with the exact SE
values under periodic boundary conditions show almost perfect
agreement, considerably better than that of experiments with the exact
SE values for an infinite system.

The friction coefficients $\stensor{K}^{\alpha\alpha}_{\text{DNS}}$ of
all the configurations are plotted in fig.~\ref{f:nonspherical_k_all}
with respect to the exact SE value for a system with the same periodic
boundary conditions $\stensor{K}^{\alpha\alpha}_{\text{SE}}$. The
results clearly show the accuracy of our method, as
$\stensor{K}^{\alpha\alpha}_{\text{DNS}} =
\stensor{K}^{\alpha\alpha}_{\text{SE}}$ within $\lesssim 2\%$.
Detailed results for the hexagonal particles are given in
fig.~\ref{f:nonspherical_k_hex}, where the three form factors
$\stensor{K}^{\alpha\alpha}$ are plotted as a function of the vertical
dimensions of the agglomerate ($\sin{\theta/2}$). The DNS results show
almost perfect agreement with the exact SE results. The difference
among the frictions coefficients and their dependence on the branching
angle is very accurately reproduced. Finally, the form factors for the
rectangular $(l\times m)\times n$ arrays are plotted in
fig.~\ref{f:nonspherical_k_rect}, as a function of increasing vertical
height $n$. As expected, the agreement with the exact results is very
good, and we are able to accurately distinguish between particles with
the same cross-sectional area $(l\times n)$.
\subsection{Chiral Structures}
\begin{figure}[ht!]
  \centering
  \includegraphics[width=0.36\textwidth]{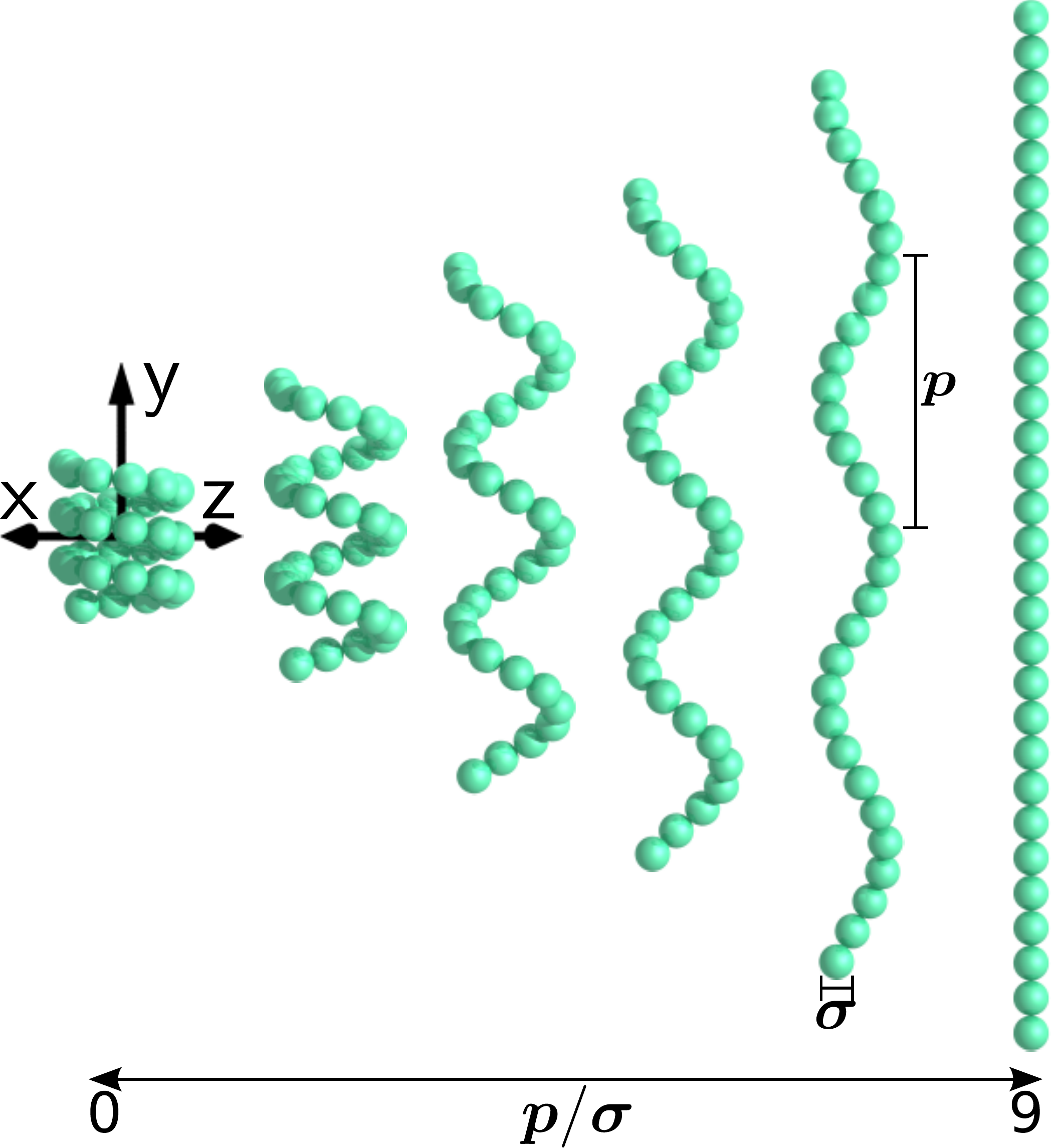}
  \caption{\label{f:helix}(color online) Helical structures of varying
    pitch $p\sigma$ ($\sigma$ is the particle diameter), with constant
    length ($30$ beads) and number of turns ($3$).}
\end{figure}
Up to now, we have considered symmetric particles for which the
translational and rotational motion are only weakly coupled, if at
all. Here we will analyze chiral structures which exhibit a very
strong coupling between the two motions. We consider left-handed
helices composed of a fixed number of beads ($30$) and turns ($3$),
which vary only in their degree of pitch $p$ (surface to surface
distance between turns). The pitch will then vary in the range $0 \le
p \le 9$, with $p=0$ corresponding to a packed configuration (hollow
cylinder), and $p=9$ to a linear chain (see fig.~\ref{f:helix}). The
simulation protocol is slightly modified with respect to the previous
systems, we now consider beads of radius $a=4$ and a simulation box
size of $L=256$. To compute the complete friction matrix, we now
require six different simulations for each helical geometry: three at
fixed velocity $V^{\alpha} = -10^{-3}$ and three at fixed angular
velocity $\Omega^{\alpha} = 10^{-4}$ ($\alpha=x,y,z$).

The friction matrices obtained from the DNS simulations, together with
the exact SE values, for a helix with pitch $p/\sigma = 6.5$ ($\sigma
= 2 a$ the diameter of the beads) are given in
eqs.~\eqref{e:helix_ktt}--\eqref{e:helix_krt} 
\begin{align}
  \tensor{K}^{tt}_{\text{DNS}} &= \left(\begin{smallmatrix}
    -3.44 & -1.38\times 10^{-2} & -7.53\times 10^{-3} \\
    -1.35\times 10^{-2} & -2.28 & -4.21\times 10^{-2} \\
    -7.44\times 10^{-3} & -4.21\times 10^{-2} & -3.47
  \end{smallmatrix}\right)\label{e:helix_ktt}\\
  \tensor{K}^{tt}_{\text{SE}\phantom{S}} &= \left(\begin{smallmatrix}
    -3.4 & -1.3\times 10^{-2} & -7.55\times 10^{-3} \\
    -1.3\times 10^{-2} & -2.22 & -4.01\times 10^{-2} \\
    -7.55\times 10^{-3} & -4.01\times 10^{-2} & -3.42 
  \end{smallmatrix}\right)\notag
\end{align}
\begin{align}
  \tensor{K}^{rr}_{\text{DNS}} &= \left(\begin{smallmatrix}
    -5.67\times 10^1 & -6.42\times 10^{-1} & 2.04\times 10^{-1} \\
    -5.59\times 10^{-1} & -2.61 & -1.62 \\
    1.56\times 10^{-1} & -1.58 & -5.6\times 10^1
  \end{smallmatrix}\right)\label{e:helix_krr}\\
  \tensor{K}^{rr}_{\text{SE}\phantom{S}} &= \left(\begin{smallmatrix}
    -5.64\times 10^1 & -5.3\times 10^{-1} & 1.55\times 10^{-1} \\
    -5.3\times 10^{-1} & -2.78 & -1.63 \\
    1.55\times 10^{-1} & -1.63 & -5.6\times 10^1    
  \end{smallmatrix}\right)\notag
\end{align}
\begin{align}
  \tensor{K}^{rt}_{\text{DNS}} &= \left(\begin{smallmatrix}
    3.94 & 6.27\times 10^{-1} & -1.19 \\
    -3.1 & -5.39 & -9.4 \\
    -6.03\times 10^{-1} & 1.94 & 1.34
  \end{smallmatrix}\right)\label{e:helix_krt}\\
  \tensor{K}^{rt}_{\text{SE}\phantom{S}} &= \left(\begin{smallmatrix}
    3.92 & 6.63\times 10^{-1} & -9.26\times 10^{-1} \\
    -2.99 & -5.21 & -9.22 \\
    -9.27\times 10^{-1} & 2.04 & 1.33    
  \end{smallmatrix}\right)\notag
\end{align}
The complex nature of the fluid flow generated by the motion of the
body is clearly evident in the form of the friction matrices. The
translational (rotational) friction matrix $\tensor{K}^{tt}$
($\tensor{K}^{rr}$) is no longer diagonal, eqs.~\eqref{e:helix_ktt}
and \eqref{e:helix_krr}, although it remains symmetric, which means
that the hydrodynamic force (torque) will not be parallel to the
direction of motion (axis of rotation). We note that although the
off-diagonal components can be up two three orders of magnitude
smaller than the diagonal components, the DNS method is able to
accurately measure all contributions. Although this accuracy is
slightly reduced when considering the coupling between translation and
rotation $\tensor{K}^{tr}$ (the small off-diagonal components can show
large relative errors $\simeq 30\%$), the dominant components are well
reproduced. To establish a clear estimate of the error, we use the
Frobenius norm $\norm{\cdot}_F$ as a measure of the difference between
the two matrices
\begin{align}
  \norm{A}_F^2 &= \sum_\alpha\sum_\beta\norm{A^{\alpha\beta}}^2
  \label{e:frobenius}
\end{align} 
The overall error of the DNS method,
computed as the relative distance between the DNS and SE friction
matrices is $\lesssim 5\%$.
\begin{align}
  \chi &= \begin{pmatrix}
    \frac{\fnorm{\tensor{K}^{tt}_{\text{SE}}-\tensor{K}^{tt}_{\text{DNS}}}}{\fnorm{\tensor{K}^{tt}_{\text{SE}}}},&
    \frac{\fnorm{\tensor{K}^{rt}_{\text{SE}}-\tensor{K}^{rt}_{\text{DNS}}}}{\fnorm{\tensor{K}^{rt}_{\text{SE}}}}, &
    \frac{\fnorm{\tensor{K}^{rr}_{\text{SE}}-\tensor{K}^{rr}_{\text{DNS}}}}{\fnorm{\tensor{K}^{rr}_{\text{SE}}}} \\
    \end{pmatrix}\notag\\
  &= \begin{pmatrix}
    1.6, & 4.3, & 4.1\times 10^{-1}
  \end{pmatrix}10^{-2} \label{e:helix_error}
\end{align}

\begin{figure}[ht!]
  \centering
  \includegraphics[width=0.48\textwidth]{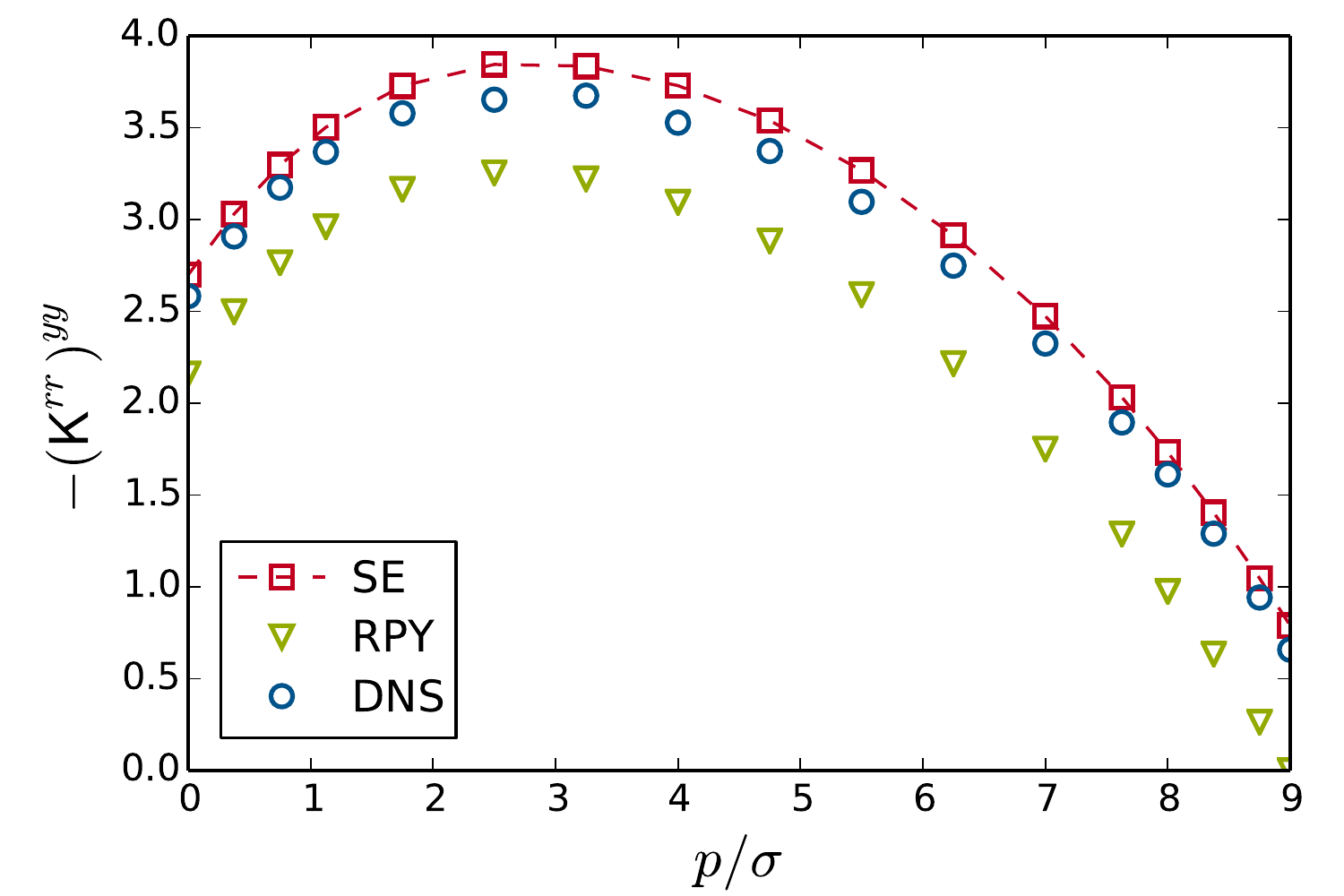}\\
  \includegraphics[width=0.48\textwidth]{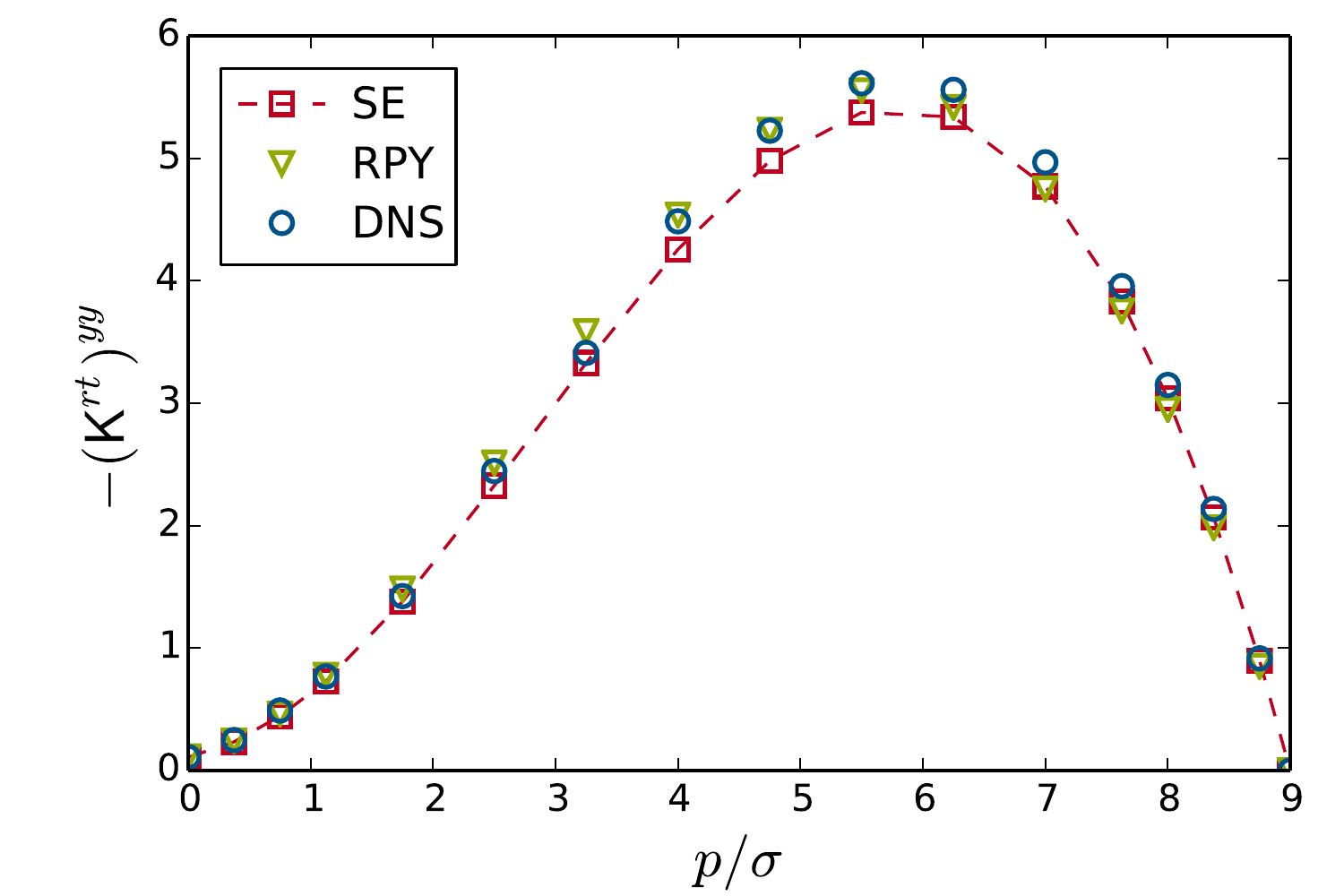}\\
  \caption{\label{f:helix_pitch} $yy$ friction coefficients for a
    helix of fixed length and number of turns, as a function of pitch
    $p/\sigma$ ($\sigma$ is the bead radius). (top) rotation-rotation
    and (bottom) rotation-translation friction tensor. The results
    obtained from DNS simulations are compared to the exact SE values,
    as well as those obtained using an RPY approximation.}
\end{figure}
Finally, to see in what way the structure of the helix will affect its
motion, we consider the $yy$ components of the $\tensor{K}^{rt}$ and
$\tensor{K}^{rr}$ friction matrices, as a function of the pitch
distance $p$. As the pitch is increased and the helix starts to
stretch, fluid flow between the turns of the helix will start to
increase, while the cross sectional area of the particle ($xz$ plane)
is reduced. The latter will increase the torque felt by the helix (as
the particle must now drag the fluid along), while the former will
tend to reduce it (by reducing the moment of inertia). These effects
give rise to the behavior seen in fig~\ref{f:helix_pitch}, where the
$\tensor{K}^{rr}$ coefficient shows a maximum at an intermediate pitch
value $p/\sigma \simeq 3$. Similar behavior is observed for the
$\tensor{K}^{rt}$ coefficient, although the maximum is obtained at a
different pitch value $p/\sigma \simeq 6$, and the coupling between
rotation and translation disappears for $p/\sigma=0$ and $p/\sigma=9$,
as expected. For comparison purposes, we have also plotted the results
obtained using the Ewald-summed RPY tensor\cite{Beenakker:1986cc} in
fig~\ref{f:helix_pitch}. The agreement is surprisingly good for the
$\tensor{K}^{tr}$ tensor, but a considerable discrepancy appears in
the coefficients of the $\tensor{K}^{rr}$ tensor, as this RPY
formulation does not include hydrodynamic effects arising from the
particle rotation.

\section{Conclusions}
We have extended the SP method to apply to arbitrary rigid bodies. For
the moment, we have considered only particles constructed as a rigid
agglomerate of (possibly overlapping) spherical beads, but alternative
formulations are straightforward. We have verified the accuracy of our
method by performing low Reynolds number DNS simulations to compute
the single particle friction coefficients for a large variety of rigid
bodies. Our results were compared with the exact solutions to the
Stokes equation, showing excellent agreement in all cases. While there
are several methods capable of performing these type of calculations,
they impose a number of restrictions which can severely limit the type
of systems to which they can be applied. Our method can capture the
many-body hydrodynamic effects very accurately, and it is not
restricted to zero Reynolds number flow or Newtonian host solvents. In
future papers we will consider the lubrication effects of
non-spherical particles (chains, rods, disks, tori, etc), as well as
the dynamics at high Reynolds number and in the presence of a
background flow fields.

\begin{acknowledgements}
  The authors would like to express their gratitude to the Japan
  Society for the Promotion of Science for financial support
  (Grants-in-Aid for Scientific Research KAKENHI no. 23244087) and Mr.
  Takuya Kobiki for his help with the simulations on helical particles.
\end{acknowledgements}

%
\end{document}